# Robustness of different modifications of Grover's algorithm based on generalized Householder reflections with different phases

*Hristo Tonchev[1], Petar Danev[1]*

[1] *Institute for Nuclear Research and Nuclear Energy, Bulgarian Academy of Sciences, 72 Tzarigradsko Chaussée, 1784 Sofia, Bulgaria*
 Emails: htonchev@ inrne.bas.bg    pdanev@inrne.bas.bg

***Abstract:*** *In this work we study five Grover's algorithm modifications, where each iteration is constructed by two generalized Householder reflections, against inaccuracies in the phases. By using semi-empirical methods, we investigate various characteristics of the dependence between the probability to find solution and the phase errors. The first of them is the robustness of the probability to errors in the phase. The second one is how quickly the probability falls beyond the stability interval. And finally, the average success rate of the algorithm when the parameters are in the range of the highly robust interval. Two of the modifications require usage of the same Grover operator each iteration and in the other three it differs. Those semi-empirical methods give us the, tool to make prediction of the quantum algorithm modifications' overall behavior and compare them for even larger register size.*

1. **Introduction**

Grover's algorithm is a quantum algorithm for searching for elements that meet certain criteria in a linear unordered database. This algorithm, which was first introduced by Lov Grover [1] in 1996, requires one quantum register of size equal to the number of searched elements and is quadratically faster than the best classical search algorithms. Together with Shor's factoring algorithm [2] they show that using quantum algorithms can solve certain problems faster than using classical algorithms [3]. This sparked interest in quantum information in general and lead to the development of various quantum algorithms. Currently, there are more than sixty quantum algorithms [4] that solve problems faster than their corresponding classical counterparts.

The original Grover's algorithm is probabilistic [3]. However, there are some modifications that can make it deterministic. Such modifications are obtained by performing a single iteration with a smaller angle [5], by changing the initial state [6] and by changing the operators [7] [8] [9]. The first way to make the latter change is to use generalized Householder reflections with the same phase to construct both operators (oracle and a reflection against state |0⟩) in Grover's iteration (phase matching) [7]. The second way to modify the operators is to use one generalized Householder's reflection and operators that are constructed from different phase multipliers (multiphase matching) [8]. In fact, a whole family of operators that can be used to construct Grover's algorithm is shown in [9]. Grover's algorithm can also be modified to achieve a different probability of finding each of the targets [10]. This modification was also improved to use phase and multiphase matches [11].

Grover's algorithm is not important only by itself in order to find elements with certain properties (minimal, maximal and so on), but it is also used as a subroutine of other algorithms like quantum counting algorithm (finds the number of solutions) [3], speeding up finding a solution to a nonpolynomial complete problem [3], Brassard-Hoyer-Tapp (BHT) algorithm for finding collision detection [12]. Grover's algorithm is also used in cryptography for speeding up an exhaustive search attack on classical cryptographic protocols [13], BHT can be used for attacking Hash functions [14], quantum key distribution [15], two types of quantum secret sharing protocol [16], [17] and other. Quantum machine learning also uses Grover's algorithm to construct quantum support vector machines (in binary classification problems) [18].

Implementations of quantum computers can be made on various physical systems exhibiting quantum behavior. Some examples of such systems and corresponding qubit implementations are: ion traps (electron levels) [19], linear optical system with single photon sources (photon qubits) [20], quantum dots (electron spin) [21], diamond with nitrogen vacancies (electron spin) [22] and superconducting qubits based on Josephson transition [23]. All physical systems have different advantages and disadvantages.

For example, by using of generalized Householder reflections, certain gates can be made very efficiently on photonic - [24] and ion trap-based [25] quantum computers. The generalized Householder reflection has one additional advantage - it can be used for faster (and more fault tolerant) quantum gates. Constructing operators by decomposition into Householder reflections [26] is quadratically faster than using Given rotations [3]. In addition, the decomposition into Householder reflections can be used in the case of a qudit of arbitrary dimension [27]. Grover's algorithm with operators constructed by generalized Householder reflections in both photonic and ion trap architectures has already been discussed in [24] and [28].

To achieve their speedup, quantum algorithms use quantum effects such as entanglement, quantum superposition, and quantum interferences. The quantum states are fragile and external interaction can change or destroy them easily – this raises the need to study ways to make the experimental implementation more reliable and to make the quantum algorithms more robust. Extensive studies of decoherence in ion traps due to heat and noise can be found in [29] and [30], respectively.

In the case of Grover's algorithm, the decoherence reduces the probability of finding a solution by increasing the number of iterations required. Often, the effect of noise on quantum systems is studied by considering imperfect gates [3] or by adding additional noise gates [31]. There is a Grover's algorithm modification, where the oracle has a small probability to not recognize some solutions at each iteration [32]. It is shown that the algorithm will become unusable if the number of errors during the execution exceeds a certain threshold [30]. Grover's algorithm was also modified to: make it more error-tolerant [33], add measurement error mitigation [34] and quantum error correction [35].

Inaccuracies in the phases can also lower the security against interception attacks in the quantum cryptographic protocol based on the Grover's algorithm [36], constructed by generalized Householder reflections. Those inaccuracies will decrease the probability that the participants will obtain the secret, while the probability that an eavesdropper is able to crack the protocol remains unchanged.

The Hill function [37], [38] is a logistical regression function that has found application in systems biology, pharmacology, mathematical and computation biology to describe nonlinear processes. The original function has similar behavior to the sigmoid function; however, it can be modified in order to describe a Gaussian-like function with a controllable slope [39]. The modified Hill function was applied to approximate and analyze the quantum random search algorithm's behavior [40].

In our previous work, we have used different methods to study the robustness against phase errors of the quantum random walk search algorithm on hypercube with traversing coins constructed by using generalized Householder reflection [41], [39]. We have also investigated the robustness of some of its modifications [42], [43]. Here, we will make a similar study of different modifications of the Grover's algorithm.

This work is structured as follows: In Section 2, a brief description of the Grover search algorithm is given along with its geometric interpretation. Next, in Section 3 two modifications of the algorithm that use the same phases in the generalized Householder reflections in all iterations are showed (Section 3.1 and Section 3.3), numerical simulations of their robustness (stability against inaccuracy in their phases) were discussed in Section 3.2 and Section 3.4. The general case of modifications that use different phases of generalized Householder reflection each iteration is described briefly in Section 4. One of our main tools used for studying and comparing the modifications of the Grover's algorithm robustness is discussed in Section 5. In Section 6, we study few different modifications of Grover's algorithm which incorporates four distinct phases. Section 7 compares the robustness of all modifications described in this work. We briefly summarize our result in the Conclusion (Section 8).

2. **Original Grover's search Algorithm**

Grover's algorithm is a probabilistic quantum algorithm that searches a linear unordered database quadratically faster than any classical algorithm. Compared to quantum random walk search, it has some advantages. It requires fewer qubits to implement (quantum walk search requires two additional registers, one control and one for the probabilities to go in each direction) and fewer gates at each iteration (it does not require control gates or a shift operator).

The algorithm finds one of $M$ solutions in a register consisting of $N$ elements. It uses a single register of dimension $N$ that stores all the states to be searched. The register can consist of qubits or qudits of arbitrary dimension. Its initial state is $|0\rangle$.

A discrete Fourier transform matrix (unitary matrix with its first column containing only elements with equal amplitudes [28]) denoted by $F$ has to be applied to the initial state to obtain an equal superposition of all states.

$$|\psi\rangle_N = F|0\rangle_N = \frac{1}{\sqrt{d}} \sum_{i=0}^{d-1} |i\rangle_N \qquad (1)$$

Here, $|i\rangle_N$ are the number states in the computational basis. The subscript denotes the dimension of the quantum state.

Next, the Grover iteration should be applied on this state a fixed number of times $k_{iter}$. Grover's iteration consists of the following operator:

$$G_N(M) = P_N O_N(M). \qquad (2)$$

Here, $O_N(M)$ is the oracle that recognizes the solutions and marks them by changing their sign. Mathematically this can be written as:

$$O_N(M) = I_N - 2|\beta\rangle_N \langle\beta|_N \qquad (3)$$

where $I_N$ is the identity matrix with dimension $N$ and $|\beta\rangle_N$ is an equal superposition of all solutions $M$.

The second gate is a reflection against the state at the beginning of the algorithm.

$$P_N = I_N - 2|\psi\rangle_N \langle\psi|_N \qquad (4)$$

In the original algorithm (in both cases for qubits and qudits), the second reflection is constructed as $P_N = F_N \, Diag(-1,1 \dots ,1)_N F_N^{-1}$.

On Fig. 1 is shown the quantum circuit of the algorithm:

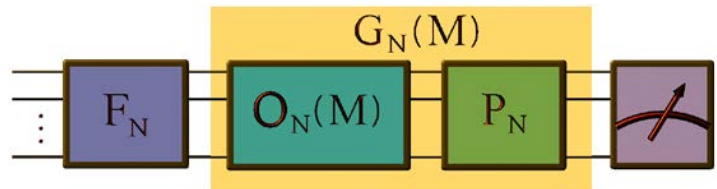

*Fig. 1. Quantum circuit of Grover's search algorithm. An iteration of the algorithm, written as $G_N$, consists of an oracle $O_N$ and a reflection against the previous state $P_N$. Discrete Fourier transform is denoted by $F_N$. The parameters N and M correspond to the register's size and the number of solutions.*

Grover's iteration is a rotation because it consists of two reflections. It is performed in the plane spanned between the vector $\beta$, that is the equal superposition of all solutions, and the vector $\alpha$, that is the equal superposition of all non-solution elements.

$$|\alpha\rangle_N = \frac{1}{\sqrt{N-M}} \sum_{\substack{i=0 \\ n_i \notin \{m_0, m_1, \ldots, m_{M-1}\}}}^{N-M+1} |n_i\rangle_N \quad (5)$$

$$|\beta\rangle_N = \frac{1}{\sqrt{M}} \sum_{i=0}^{M} |m_i\rangle_N \quad (6)$$

The Grover's algorithm can be represented geometrically in the following way:

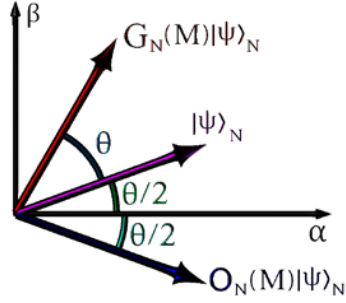

Fig. 2. Geometric representation of the Grover's algorithm. The oracle $O_N$ rotates the wave function $|\psi\rangle$ on an angle $\theta/2$ against the axis of all not solutions $\alpha$. Completing the Grover's iteration, by applying the operator P, makes a second rotation against $|\psi\rangle$ on angle $\theta$. The final goal is to rotate the vector $|\psi\rangle$ as close as possible to the axis of all solutions $\beta$.

The initial state can be written in the basis of those vectors as:

$$|\psi\rangle_N = \cos\left(\frac{\theta}{2}\right)|\alpha\rangle_N + \sin\left(\frac{\theta}{2}\right)|\beta\rangle_N \quad (7)$$

Each iteration leads to rotation of the wave function in this plane on an angle $\theta$:

$$\theta = 2\arcsin(\sqrt{M/N}) \quad (8)$$

After $k_{iter}$ iterations the state becomes:

$$(G_N(M))^{k_{iter}}|\psi\rangle_N = \cos\left(\frac{2k_{iter}+1}{2}\theta\right)|\alpha\rangle_N + \sin\left(\frac{2k_{iter}+1}{2}\theta\right)|\beta\rangle_N \quad (9)$$

In order to find the number of iterations required to maximize the probability to find the solution, $\cos((2k_{iter}+1)\theta/2)$ should be as close as possible to zero:

$$\frac{\pi}{2} \cong \left(k_{iter} + \frac{1}{2}\right)\theta \quad (10)$$

The number of iterations should always be an integer. This means that, the required number of iterations is the smallest integer $k_{iter}$ that fulfills the above equation. It can be easily calculated by using the register's size:

$$k_{iter} = \left\lceil \frac{\pi}{4}\sqrt{N/M} \right\rceil \quad (11)$$

and $\lceil \ \rceil$ represents rounding up the content in the brackets.

### 3. Grover's algorithm with generalized Householder reflections

Here, we will show two modifications of the algorithm (shown in [7] and [44]) that use generalized Householder reflection.

#### 3.1. First phase matching modification

In order to obtain probability to find solution equal to one, in the construction of the operators instead of Householder reflection, a generalized Householder reflection can be used.

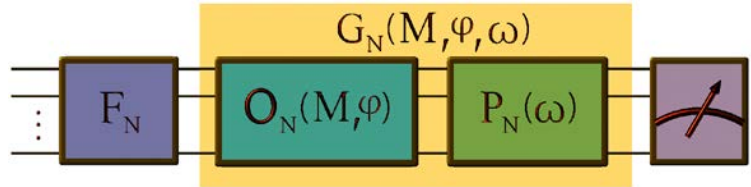

Fig. 3. *Quantum circuit of Grover's search algorithm modified to have zero failure rate. Householder reflections $O_N(M)$ and $P_N$ in the original algorithm are changed by generalized Householder reflections $O_N(M, \varphi)$ and $P_N(\omega)$*

The oracle and the reflection against the initial state becomes:

$$O(M, \varphi) = I_N - (1 - e^{i\varphi})|\beta\rangle_N \langle\beta|_N \tag{12}$$

$$P_N(\omega) = I_N - (1 - e^{i\omega})|\psi\rangle_N \langle\psi|_N \tag{13}$$

Let us denote the Grover's iteration in this case as $G_N^+$:

$$G_N(M, \varphi, \omega) = P_N(\omega)O(M, \varphi) = G_N^+ \tag{14}$$

Using a generalized Householder reflection allows smaller angle rotations to be made and thus achieve

$$\cos\left(\frac{(2k_{iter} + 1)\theta}{2}\right) = 0 \tag{15}$$

To obtain a probability of finding a solution equal to one, it has already been shown that [7] the angles φ and ω must be the same and their value have to be:

$$\varphi_{max} = \omega_{max} = 2\arcsin\left(\sin\left(\frac{\pi}{4J_\omega + 6}\right)\sqrt{N}\right) \tag{16}$$

where $J$ is a substitution of the following expression:

$$J_\omega = \left\lfloor \frac{|0.5\pi - \theta|}{2\theta} \right\rfloor \tag{17}$$

and $\lfloor \ \rfloor$ represents rounding down the content in the brackets.

It is also important to note that together with $\varphi$ that achieves probability to find a solution equal to 1, the angle $2\pi - \varphi$ also achieves the same probability.

### 3.2. Robustness of the first modification that uses phase matching ($OPH$)

In case of arbitrary phases of both Householder reflections (in O and P operators), the probability of the Grover's algorithm to find a solution after $r$ iterations can easily be calculated recursively by using Eq. (9) and Eq. (11) together with the following equations:

$$|\psi(r)\rangle_N = G^r |\psi\rangle_N \tag{18}$$

$$G|\beta\rangle_N = \frac{e^{i\varphi}(e^{i\omega} - 1)}{2} \sin(\theta) |\alpha\rangle_N + e^{i\varphi}\left(1 - \frac{(e^{i\omega} - 1)}{2}(\cos(\theta) - 1)\right)|\beta\rangle_N \tag{19}$$

$$G|\alpha\rangle_N = 1 + \frac{(e^{i\omega} - 1)}{2}(1 + \cos(\theta))|\alpha\rangle_N + \frac{(e^{i\omega} - 1)}{2}\sin(\theta)|\beta\rangle_N \tag{20}$$

where $\theta$ and $|\psi\rangle_N$ are defined by Eq. (15) and Eq. (7).

For example, let there be only one solution $|m\rangle_N$. The probability to find it in case of one iteration is shown below (it should be used for register sizes up to 6 included):

$$\begin{aligned} p(m) &= \langle m|_N G|\psi\rangle_N \\ &= e^{i\varphi} \sin\left(\frac{\theta}{2}\right)\left(1 + (e^{i\omega} - 1)\sin^2\left(\frac{\theta}{2}\right)\right) \\ &+ \frac{1}{2}(e^{i\omega} - 1)\cos\left(\frac{\theta}{2}\right)\sin(\theta)|\beta\rangle_N \end{aligned} \tag{21}$$

The probability to find solution after two iterations is (it should be used for register sizes between 7 and 14):

$$\begin{aligned} p(m) &= \langle m|_N GG|\psi\rangle_N \\ &= \left(e^{i\varphi}\sin\left(\frac{\theta}{2}\right)\left(1 + (e^{i\omega} - 1)\sin\left(\frac{\theta}{2}\right)^2\right)\left(e^{i\varphi} + e^{i\omega} - 1 \right.\right. \\ &\left. + (e^{i\varphi} - 1)(e^{i\omega} - 1)\sin^2\left(\frac{\theta}{2}\right)\right) \\ &+ \frac{1}{2}(e^{i\omega} - 1)\cos\left(\frac{\theta}{2}\right)\left(e^{i\omega} \right. \\ &\left.\left. + (e^{i\varphi} - 1)(e^{i\omega} - 1)\sin^2\left(\frac{\theta}{2}\right)\right)\sin(\theta)\right) \end{aligned} \tag{22}$$

On Fig. *4* are shown as an example the numerical simulations of the probability to find solution that depends on both Householder phases φ and ω for a fixed register size and when only one element is a solution.

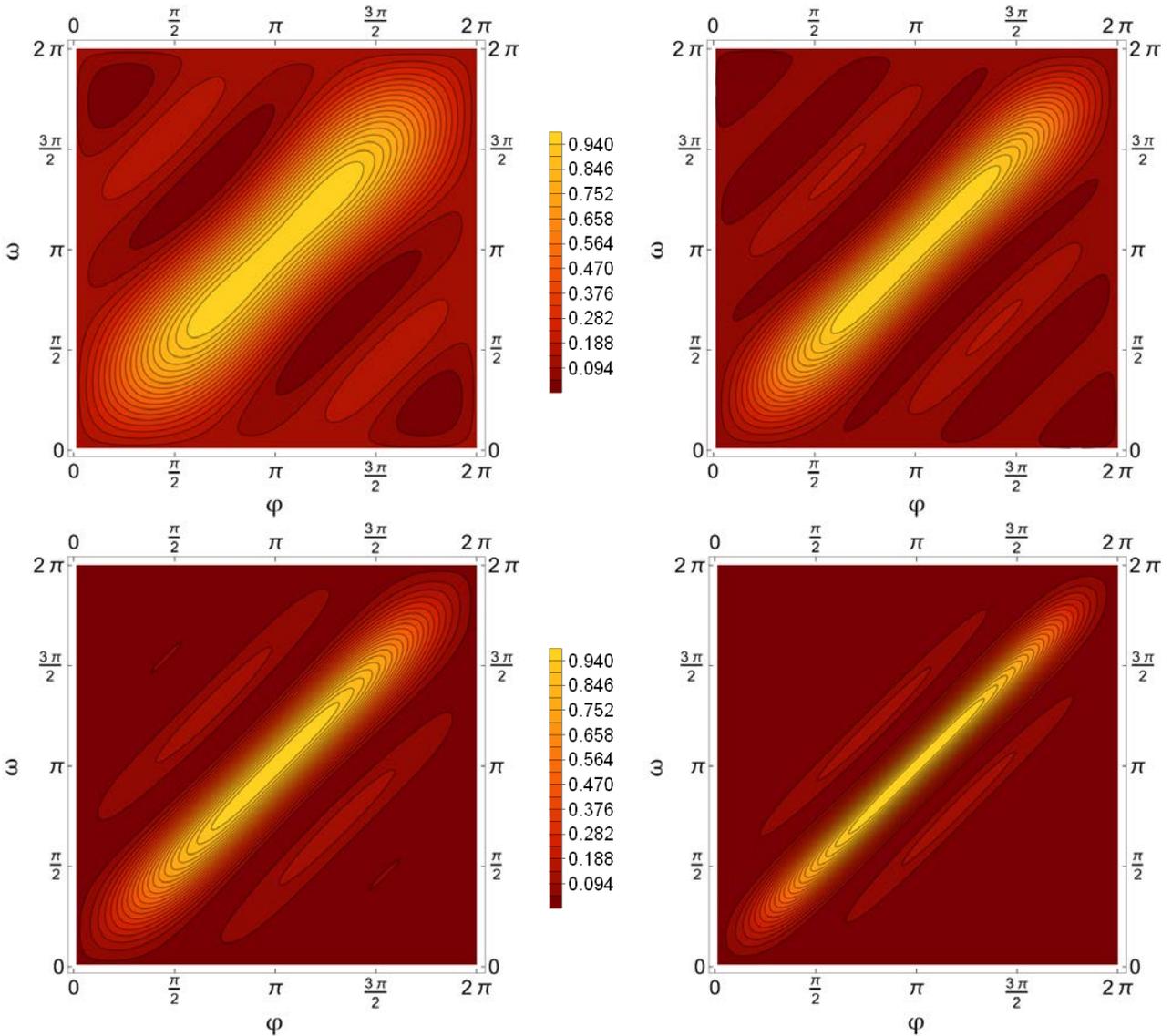

*Fig. 4. Probability to find solution depending on both angles (φ, ω on vertical and horizontal axes respectively) used in the Grover's algorithm with its iteration constructed by using generalized Householder reflection. The highest probability to find solution is shown in yellow and the lowest – in red. The top left and right pictures correspond to register sizes 9 and 18 respectively, and the bottom left and right correspond to register sizes 36 and 72 respectively.*

If we make a cross-section passing through the point $\{\varphi = \pi, \omega = \pi\}$, regardless of the angle between it and the coordinate axes, the probability of finding a solution will be described by a Gaussian-like curve. We will study four relations between the phases.

In the first group, the probability to find solution can be written as:

$$p(m, \varphi, \omega) \rightarrow p(m = const, \varphi, \omega(\varphi)) \equiv p(\varphi) \tag{23}$$

The relation above can be used to study the following functional dependences:

$$\omega(\varphi) = \pi \tag{24}$$

$$\omega(\varphi) = \varphi \tag{25}$$

$$\omega(\varphi) = 2\pi - \varphi \tag{26}$$

$$\varphi(\omega) = \pi \tag{27}$$

It is important to note that when $\varphi = \pi$, p depends only on ω:

$$p(m, \varphi, \omega) \rightarrow p(m = const, \varphi(\omega), \omega) \equiv p(\omega) \tag{28}$$

Thus, further in the text the probability $p$ in the case of Eq. (27) will be written as $p(\omega)$.

Those functional dependences between phases (Eqs. (24), (25), (26) and (27)) are chosen because they define the directions in the plane defined by $\varphi$ and $\omega$ in which the modifications to be studied have the best or the worst performance. This would allow us to compare the robustness of the different modifications of the Grover's algorithm. In case of fixed relation between phases we can use semiempirical methods to obtain different characteristics of the probability to find solution against errors in the phases (more will be explained later in this chapter). When we compare the stability of the algorithm in case of each of those relations (Eqs. (24), (25), (26) and (27)) for different modifications, we can estimate the robustness in the best and worst cases. In four of those modifications, the most and the least robust cases correspond to two of those relations exactly. In the fifth case these are very good approximation for large register.

Analytical formulas for probability to find solution for register size 9 and different $\omega$ are shown below:

$$p(\varphi, \omega = \varphi) = |0.03292 + 0.46090 e^{i\varphi} - 0.69135 e^{2i\varphi} \\ - 0.13168 e^{3i\varphi} - 0.00411 e^{4i\varphi}|^2 \tag{29}$$

$$p(\varphi, \omega = 2\pi - \varphi) \\ = |0.13580 - 0.32921.\cos(\varphi) + 0.52674.\cos(2\varphi)|^2 \tag{30}$$

$$p(\varphi, \omega = \pi) = |0.46090 - 0.32921 e^{i\varphi} + 0.20164 e^{2i\varphi}|^2 \tag{31}$$

$$p(\varphi = \pi, \omega) = |0.46090 - 0.32921 e^{i\omega} + 0.20164 e^{2i\omega}|^2 \tag{32}$$

Numerical simulations for few register sizes and angles are shown on *Fig. 5*. Different colour of the line corresponds to a different functional dependence between the angles. The dotted red to $\omega = 2\pi - \varphi$, the dot-dashed purple to $\varphi = \pi$, the solid teal to $\omega = \pi$ and the dashed green to $\omega = \varphi$. The upper left and right pictures correspond to register sizes 9 and 18, respectively, and the lower left and right correspond to register sizes 36 and 72, respectively. For functional dependences $\omega = \varphi$, $\omega = \pi$, and $\omega = 2\pi - \varphi$ on horizontal axis is $\varphi$. In case of $\varphi = \pi$ on the horizontal axis is $\omega$ instead.

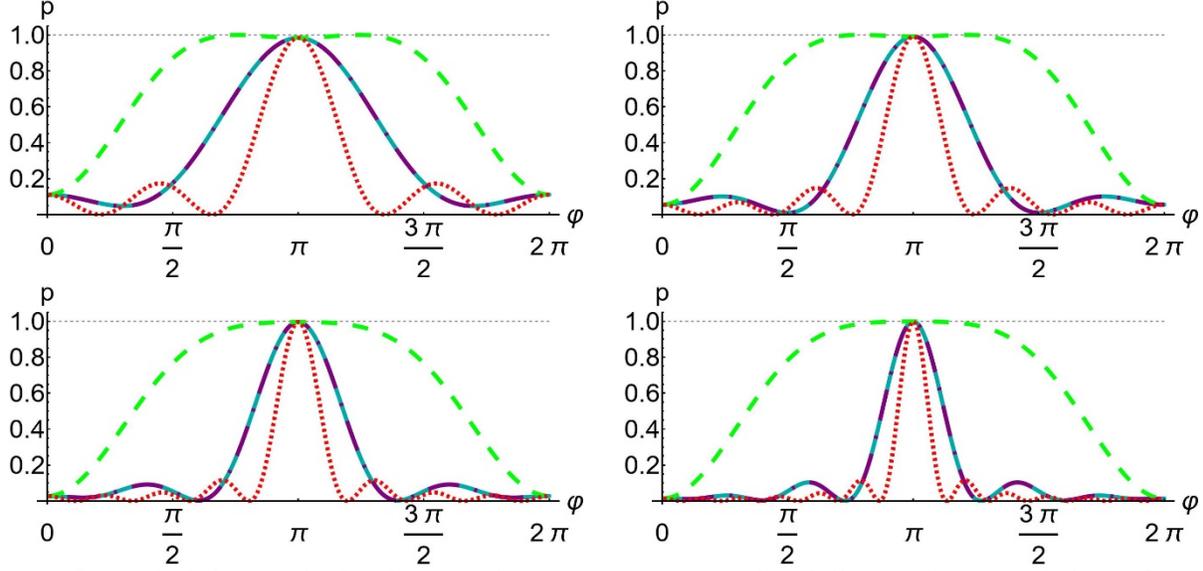

*Fig. 5. Probability to find solution after $k_{iter}$ iterations for different functional dependences between the angles used in the construction of Grover's iteration. The presented simulation results are in the case of phase matching modification and for different values of the angle $\varphi$ ($\omega$ in case of $\varphi = \pi$)(horizontal axis). The dotted red line corresponds to $\omega = 2\pi - \varphi$, the solid teal line to $\omega = \pi$, the purple dot dashed line to $\varphi = \pi$, and the dashed green line to $\omega = \varphi$. The top left, top right, bottom left and bottom right pictures correspond to register sizes 9, 18, 36 and 72 respectively.*

In this work we will study the errors in the angles $\omega$ and $\varphi$. It can be seen that the algorithm is most stable against phase errors when $\omega = \varphi$ and most unstable when $\omega = 2\pi - \varphi$. We will investigate the robustness of Grover's algorithm - robustness against inaccuracies in the two Householder phases $\varphi$, $\omega$. We define the parameters $\varepsilon_\varphi^-, \varepsilon_\varphi^+, \varepsilon_\omega^-$ and $\varepsilon_\omega^+$ which correspond to the errors in the respective phases:

$$\left(\varphi \in \left(\varphi_{max} - \varepsilon_\varphi^-, \varphi_{max} + \varepsilon_\varphi^+\right)\right) \cong p_{max}(\varphi_{max}, \omega = \text{const}, \theta = \text{const}) \equiv p(\varphi_{max}) \quad (33)$$

$$\left(\omega \in \left(\omega_{max} - \varepsilon_\omega^-, \omega_{max} + \varepsilon_\omega^+\right)\right) \cong p_{max}(\omega_{max}, \varphi = \text{const}, \theta = \text{const}) \equiv p(\omega_{max}) \quad (34)$$

In order to achieve maximum stability against $\varphi$, while having an equal probability of the error being in both directions, the relations $\varepsilon_\varphi^- = \varepsilon_\varphi^+ = \varepsilon_\varphi$ and $\varepsilon_\omega^- = \varepsilon_\omega^+ = \varepsilon_\omega$ must hold. The maximal robustness against the angle $\varphi$ (or $\omega$) is achieved when $\varepsilon_\varphi$ (or $\varepsilon_\omega$) have maximal width for $p(\varphi \in (\varphi_{max} - \varepsilon_\varphi, \varphi_{max} + \varepsilon_\varphi)) \cong p_{max}$.

In case of large register $\varphi_{max} \to \pi$. For this reason, we will focus on the robustness around the point $\{\varphi = \pi, \omega = \pi\}$.

### 3.3. Description of a second modification that uses phase matching

Another way to construct a Grover's algorithm with probability to find solution equal to one is to replace both reflections with the following operators as shown in [44]:

$$O_N(M, \varphi) = I_N - (1 - e^{i\varphi})|\beta\rangle_N\langle\beta|_N \qquad (35)$$

$$P'_N(\omega) = I_N e^{i\omega} + (1 - e^{i\omega})|\beta\rangle_N\langle\beta|_N \qquad (36)$$

The second gate can also be constructed (up to general phase multiplier) by using a generalized Householder reflection:

$$P'_N(\omega) = e^{i\omega}(I_N - (1 - e^{-i\omega})|\beta\rangle_N\langle\beta|_N) \cong P_N(-\omega) \qquad (37)$$

An iteration constructed by those two operators rotate the state in the space defined by $|\alpha\rangle_N$ and $|\beta\rangle_N$ on the same angle as in the phase modified algorithm explained in Sec. 3.1. Let us denote the Grover's iteration in this case as $G_N^-$:

$$G_N(M, \varphi, -\omega) = P_N(-\omega)O(M, \varphi) = G_N^- \qquad (38)$$

In order to obtain the probability $p$, it has already been shown that [44] the angles $\varphi$ and $\omega$ should be equal, and their value should be the same as in the case of the phase matching modification:

$$\varphi_{max} = 2\arcsin\left(\sin\left(\frac{\pi}{4J_\omega + 6}\right)\sqrt{N}\right) \qquad (39)$$

$$\omega_{max} = 2\pi - \varphi_{max} \qquad (40)$$

In is also important to note that both $\varphi$ and $2\pi - \varphi$ achieve probability to find solution equal to 1.

### 3.4. Robustness when the second type of phase matching modification is used

Similarly, to the investigations in Sec. 3.2, when both phases are arbitrary, here we can recursively obtain the probability to find solution after selected number of iterations by equations (15), (7), (18) and simply substituting $e^{i\omega}$ with $e^{-i\omega}$ in equations (19) and (20). For example, in case of two iterations (for register sizes between 7 and 14 included) and one solution, the probability to find solution is:

$$p(m) = \langle m|_N GG|\psi\rangle_N$$

$$
\begin{aligned}
= \Bigg( & e^{i\varphi}\sin\left(\frac{\theta}{2}\right)\left(1 + (e^{-i\omega} - 1)\sin\left(\frac{\theta}{2}\right)^2\right)\left(e^{i\varphi} + e^{-i\omega} - 1\right. \\
& + \left(e^{i\varphi} - 1\right)\left(e^{-i\omega} - 1\right)\sin^2\left(\frac{\theta}{2}\right)\Bigg) \\
& + \frac{1}{2}(e^{-i\omega} - 1)\cos\left(\frac{\theta}{2}\right)\Bigg(e^{-i\omega} \\
& + \left(e^{i\varphi} - 1\right)\left(e^{-i\omega} - 1\right)\sin^2\left(\frac{\theta}{2}\right)\Bigg)\sin(\theta) \Bigg)
\end{aligned}
\tag{41}
$$

Again, if we have linear functional dependence between phases that pass through $\{\varphi = \pi, \omega = \pi\}$, the probability to find solutions can be approximated with a Gaussian like curve. This observation will be important later in this work.

On *Fig. 6* are shown numerical simulations of the probability to find solution as a function of both phases $\varphi$ and $\omega$. Each picture corresponds to different register size:

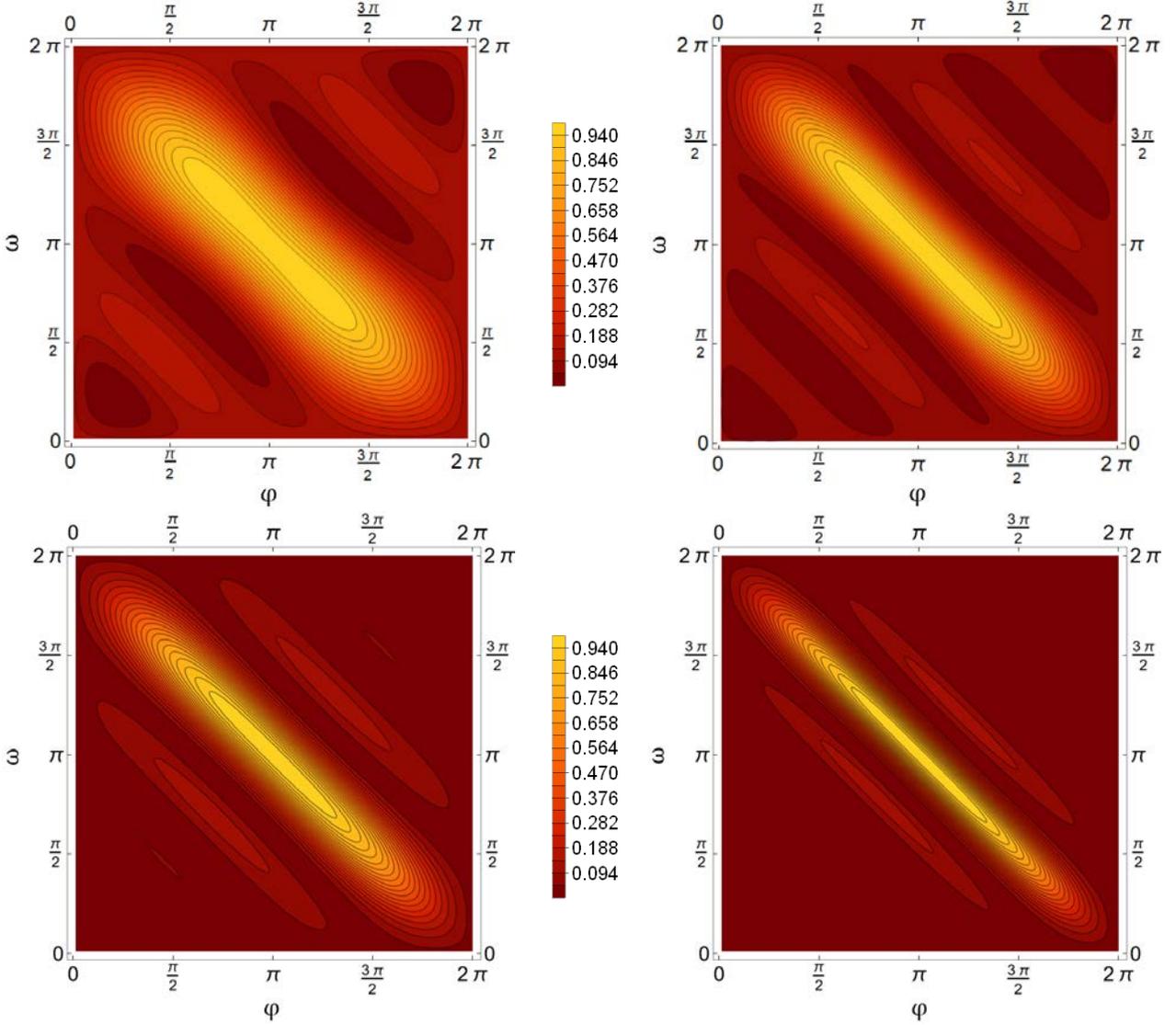

*Fig. 6. Probability to find solution after $k_{iter}$ iterations depending on both angles used to construct generalised Householder reflections in the Grover's iteration. The angles φ and ω are given on the vertical and horizontal axes, respectively. Lighter colour corresponds to higher value of p. The top left, top right, bottom left and bottom right pictures show results for registers with sizes 9, 18, 36, and 72 correspondingly.*

The analytical formula (41) giving the probability to find solution for register size 9 and angles $\omega = \pi$ or $\varphi = \pi$ coincides with formula (22) obtained by the modification discussed in 3.1. with $\omega = \pi$ or $\varphi = \pi$. However, for $\omega = \varphi$ formula (41) coincides with the one for $\omega = 2\pi - \varphi$ in the phase matching case (22). Similarly, formula for $\omega = 2\pi - \varphi$ coincides with the corresponding formula (41) when $\omega = \phi$ in the phase matching case (22). Those relations are preserved for higher number of iterations too, as can be seen by comparing the numerical simulations shown on *Fig. 4* and *Fig. 6*.

Numerical simulations of $p$ for different functional dependences between the generalized Householder phases and for four selected coin size dimensions are shown on *Fig. 7*. The relations between the angles are represented by different colours: the dotted red line is obtained with $\omega = 2\pi - \varphi$, the solid teal - with $\omega = \pi$, the dot-dashed purple - with $\varphi = \pi$ and the dashed green - with $\omega = \varphi$. The horizontal axis corresponds to angle $\phi$ in all cases except $\phi = \pi$, where the horizontal axis is given $\omega$. The top left and right and the bottom left and right pictures corespond to register sizes 9, 18, 36 and 72 respectively.

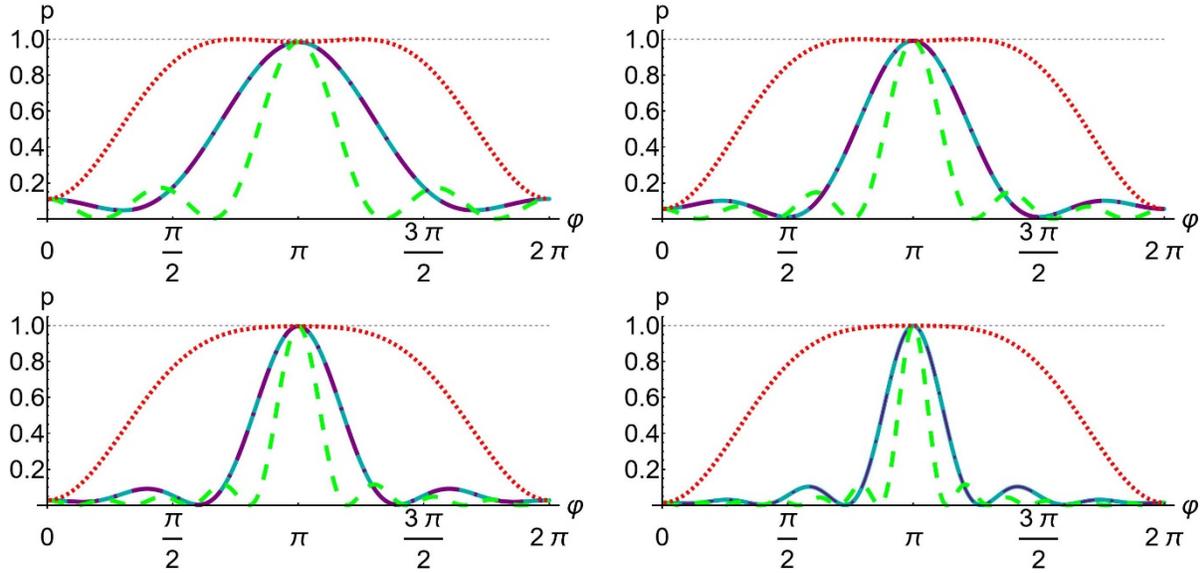

*Fig. 7. Probability to find solutions after the required number of iterations. Each line corresponds to different functional dependence between angles used in constructions of the two operators taking part in Grover's iteration - the red dotted line corresponds to $\omega = 2\pi - \varphi$, solid teal line to $\omega = \pi$, dot-dashed purple line to $\varphi = \pi$ and dashed green line to $\omega = \varphi$. The horizontal axis corresponds to angle $\omega$ in case of $\varphi = \pi$, and angle $\varphi$ in all other cases. The top left, top right, bottom left and bottom right pictures show the probability to find solution for register sizes 9, 18, 36 and 72 respectively.*

In this case, the stability of the algorithm shows the opposite behavior – it is most stable when $\omega = 2\pi - \phi$, and most unstable when $\omega = \phi$.

### 4. Grover's algorithm with multiple phases

In [8] has been studied a modification that mixes the two phase matching modifications explained before. In this modification of the algorithm, each Grover's iteration is constructed as:

$$G_N(M, \varphi_j, \omega_j) = P'_N(\omega_j) O(M, \varphi_j) \qquad (42)$$

Here, each $\varphi_j$ and $\omega_j$ in the j-th iteration can be arbitrary chosen.

$$|\psi(r)\rangle_N = \prod_{j=1}^{r} G_N(M, \varphi_j, \omega_j) |\psi\rangle_N \qquad (43)$$

The probability to find solution and the required number of iterations to obtain maximal probability depends on the register size, the number of solution and the phases used.

In this case when there is only one solution, an arbitrary register size and two iterations, the probability to find solution is the following function of both angles $\varphi_j$ and $\omega_j$:

$$p(m) = \langle m|_N G_N(M, \varphi_2, \omega_2) G_N(M, \varphi_1, \omega_1) |\psi\rangle_N =$$
$$e^{i\varphi_2} \sin\left(\frac{\theta}{2}\right) \left(-1 + e^{i\varphi_1} + e^{i\omega_1} + e_{\varphi_1} e_{\omega_1} \sin^2\left(\frac{\theta}{2}\right)\right) \qquad (44)$$
$$\left(1 + e_{\omega_2} \sin^2\left(\frac{\theta}{2}\right)\right) + \frac{1}{2} e_{\omega_2} \cos\left(\frac{\theta}{2}\right) \left(e^{i\omega_1} + e_{\varphi_1} e_{\omega_1} \sin^2\left(\frac{\theta}{2}\right)\right) \sin(\theta)$$

where the following substitutions were made: $e_{\omega_1} = (e^{i\omega_1} - 1)$, $e_{\omega_2} = (e^{i\omega_2} - 1)$, $e_{\varphi_1} = (e^{i\varphi_1} - 1)$ and $e_{\varphi_2} = (e^{i\varphi_2} - 1)$.

This modification requires the same number of iterations to find solution as in the original Grover's algorithm.

In their work Toyama et al. have studied in detail how the probability to find solutions (one or more) depends on $M/N$, when the following dependence between the phases in the generalized Householder reflection is used:

$$\omega_j = \varphi_{k_{iter}-j+1} \qquad (45)$$

The authors in [8] make analytical calculations for arbitrary phases $\varphi_j$ and show that this modification can be used to obtain high probability to find solution (equal or more than 0.998) and their solution is robust against small errors of the phase. In their paper they calculate particular examples of their modification of the Grover's algorithm, with up to 6 Grover's iterations. Particular angles mentioned in their work are the ones for which $|\varphi_j| = \pi$ for all j, and they differ only by a sign.

In this work we will use semiempirical methods to study the stability against phase errors of three similar modifications, which include multiplier to the phases that depend on the current iteration:

$$\{\varphi_j, \omega_j\} = \{\pi, (-1)^j \pi\} \qquad (46)$$

$$\{\varphi_j, \omega_j\} = \{(-1)^{j+1}\pi, (-1)^j \pi\} \qquad (47)$$

$$\{\varphi_j, \omega_j\} = \{(-1)^{\lceil j/j_{max}\rceil}\varphi, (-1)^{\lceil j/j_{max}\rceil}\omega\} \qquad (48)$$

Their robustness will also be compared to the robustness of the phase matching modification. Different characteristics of those examples will be studied in detail in the next sections.

It is also important to mention that the modification (47) is equal to the one described by (45) in the cases when the required number of iterations is an even number and in the j-th iteration is used the phase $\varphi_j = (-1)^{j+1}\pi$. In case when the required number of iterations is odd, they differ.

5. **Semi empirical method for evaluating the robustness**

The Hill function is used in to pharmacology, biochemistry, mathematical and computational biology in order to semi-empirically describe nonlinear processes of cooperative binding between one big and few small molecules. In our previous works we made a modification of this function in order to describe a gaussian and rectangular like shaped functions (See *Fig. 8*):

$$W(\varphi, b, k, n, c) = \frac{bk^n}{|\varphi - c|^n + k^n} \qquad (49)$$

Here, each of the parameters corresponds to different properties of the curve. The maximal height is controlled by b (*Fig. 8*a), the curvature and slope of the curve are controlled by n (*Fig. 8*b), the width of the bell is controlled by k (*Fig. 8*c) and the center of the bell - by c (*Fig. 8*d). Maximal height is achieved when $\varphi = c$ and in order to obtain correct results, parameters should be positive: $b, k, n > 0$.

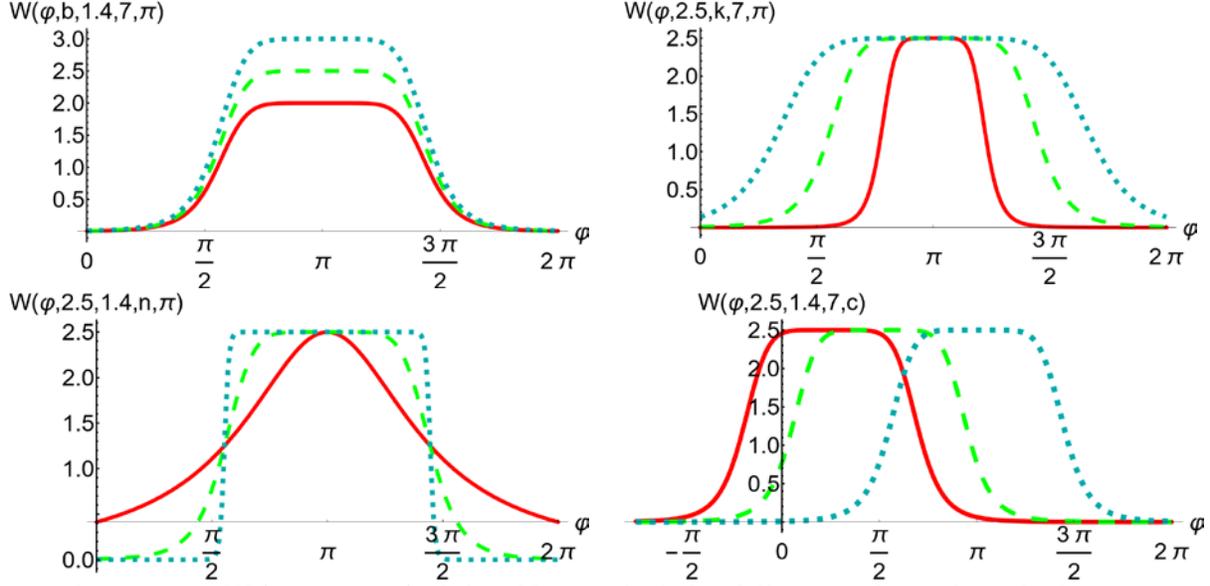

*Fig. 8. Function $W(\varphi, b, k, n, c)$ defined by Eq. (49) for different values of $\varphi$. The lines on each figure are obtained with the same values of b, k, n, and c except for one of these parameters. The top left picture corresponds to $W(\varphi, b, 1.4, 7, \pi)$ for b equal to 2 (solid red line), 2.5 (dashed green line) and 3 (dotted teal line). The top right corresponds to $W(\varphi, 2.5, k, 7, \pi)$ for k equal to 0.7 (solid line), 1.4 (dashed green line) and 2.1 (dotted teal line). The bottom left corresponds to $W(\varphi, 2.5, 1.4, n, \pi)$ for n equal to 2 (solid red line), 7 (dashed green line) and 45 (dotted teal line). The bottom right corresponds to $W(\varphi, 2.5, 1.4, c, \pi)$ for c equal to $\pi/4$ (solid red line), $\pi/2$ (dashed green line) and $\pi$ (dotted teal line).*

This property, that each of modified Hill function's parameters controls one of the characteristics of the curve (height, slope, and so on), makes this function very suitable for fitting and analyzing other more complex functions. In addition, the parameters of this fit give us quantitative characteristics for comparing different bell like curves.

In order to assess how good is the fit by the modified Hill function, we will compute its standard deviation. In the next section we will use this function to compare the probability to find solution of Grover's search algorithm for the modifications investigated above:

$$\sigma = \sqrt{\sum_{j=1}^{N} \frac{\left(W_j(\varphi_j, b, k, n, c) - p_j(\varphi_j, \omega(\varphi_j))\right)^2}{N - q}} \qquad (50)$$

Here, $p_j$ is the probability to find solutions at angle $\varphi_j$ for functional dependence between phases $\omega(\varphi_j)$ and $W_j(\varphi_j, b, k, n, c)$ is the fit by the modified Hill function with fitting parameters $b, k, n$ and $c$.

## 6. Examples for Hill fit on the probability to find solution

Here we will present fits with the modified Hill function for the reviewed (in Section 4) modifications of the Grover's algorithm for different functional dependences between the phases in the algorithm's operators. These investigations will allow us to give predictions of modifications' robustness for larger N. We will also use them to compare the robustness of the respective modifications.

### 6.1. First and second phase matching modification

On *Fig. 9* is shown a Hill fit of the first phase matching modification in case of register size 36 and only one solution. The green solid line on the top left picture corresponds to dependence $\omega = \varphi$, the red solid line on top right corresponds to $\omega = 2\pi - \varphi$, and the teal line on bottom corresponds to $\omega = \pi$. On each picture with dashed blue line is shown its corresponding Hill fit. The parameters of the fit are written in *Table 1*.

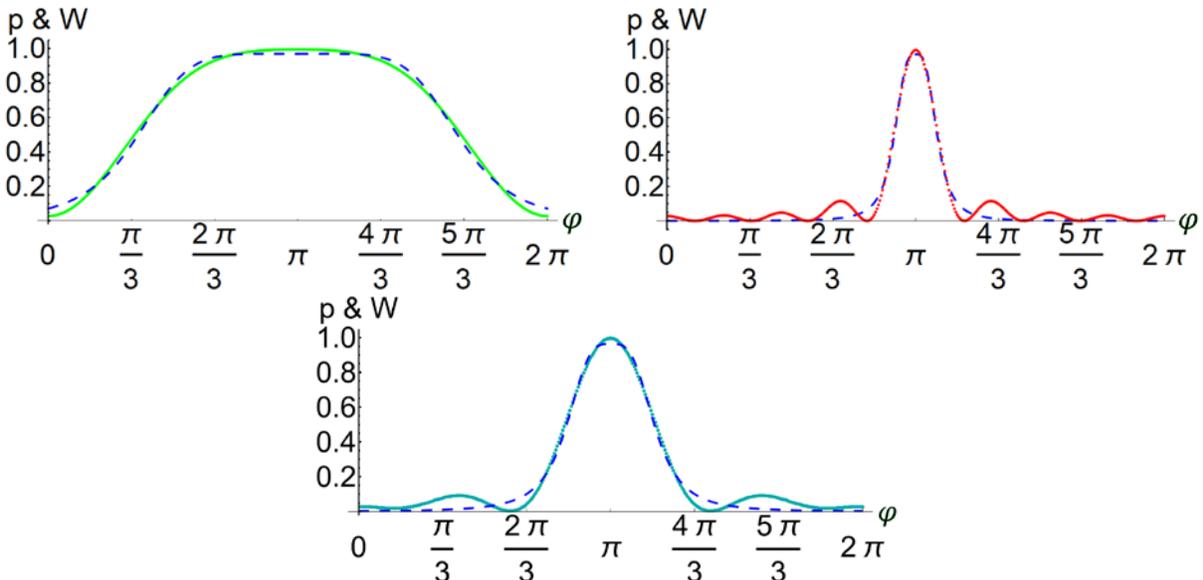

*Fig. 9. Fits with the modified Hill function of the probability to find solution after $k_{iter}$ iterations for register size 36 and different functional dependences between the phases. The solid line corresponds to the probability and the dashed line - to its fit. The top left figure corresponds to $\omega = \varphi$, the top right - to $\omega = 2\pi - \varphi$ and down is a picture for $\omega = \pi$.*

In *Table 1* are written the parameters of the modified Hill fits of the probability to find solution for different register sizes and functional dependences between the generalized Householder phases. Results are shown for register sizes {9, 36, 72 and 104} and functional dependences $\omega = \varphi$, $\omega = 2\pi - \varphi$ and $\omega = \pi$. The register size is written on the second column, the relation between the phases - on the third, the parameters b, k and n together with the standard deviation of the fit are shown on fourth, fifth, sixth and seventh column respectively.

| № | N | Dependence | b | k | n | σ |
|---|---|---|---|---|---|---|
| 1 | | $\omega = \varphi$ | 0.99162 | 2.21657 | 6.08517 | 0.00927713 |
| 2 | 9 | $\omega = 2\pi - \varphi$ | 0.988603 | 0.475057 | 2.72101 | 0.0733959 |
| 3 | | $\omega = \pi$ | 0.957434 | 1.02292 | 3.16995 | 0.0347275 |
| 4 | | $\omega = \varphi$ | 0.970608 | 2.03089 | 5.81106 | 0.0275356 |
| 5 | 36 | $\omega = 2\pi - \varphi$ | 0.970676 | 0.275992 | 3.27181 | 0.0379211 |
| 6 | | $\omega = \pi$ | 0.963316 | 0.557972 | 3.4133 | 0.0406315 |
| 7 | | $\omega = \varphi$ | 0.974974 | 2.04358 | 6.1420 | 0.0292556 |
| 8 | 72 | $\omega = 2\pi - \varphi$ | 0.972984 | 0.189477 | 3.2802 | 0.0310133 |
| 9 | | $\omega = \pi$ | 0.968527 | 0.381387 | 3.3689 | 0.0380029 |
| 10 | | $\omega = \varphi$ | 0.985716 | 2.1367 | 7.15433 | 0.0251211 |
| 11 | 104 | $\omega = 2\pi - \varphi$ | 0.975118 | 0.140599 | 3.12711 | 0.0303656 |
| 12 | | $\omega = \pi$ | 0.969398 | 0.28186 | 3.18194 | 0.0399776 |

*Table 1. Values of the parameters b, k, n and the standard deviation of the fit with the modified Hill function for different functional dependences between phases and different register size. In the second and third columns are given the register size and the functional dependence between the phases. The fourth, fifth and sixth columns show the corresponding parameters of the Hill fit and the last column shows the standard deviation of the fit.*

Those fits can be calculated for different register sizes in order to obtain an estimate of the dependence between the probability to find solution $p(N, \varphi, \omega(\varphi))$ and the Hill function parameters b(N), k(N) and n(N). On *Fig. 10.* are shown fits for N ∈ (1,111). The dashed green, solid teal, dot-dashed purple and dotted red lines represent relations $\omega = \varphi$, $\omega = 0$, $\varphi = 0$, and $\omega = 2\pi - \varphi$ respectively. The dashed blue line shows the values of the register size where the required number of Grover's iteration increases.

It can be seen on *Fig. 10.* that the standard deviation has a value between 0.12 and 0.01. It varies with increasing the register size and has fluctuations in the interval that requires the same number of iterations in order to obtain the highest probability to find solution. The standard deviation mentioned before decreases with increasing the number of iterations needed and stabilizes to around 0.04. This means that approximation of the probability to find solution for those dependences between angles with modified Hill function is very good and does not increase for large N.

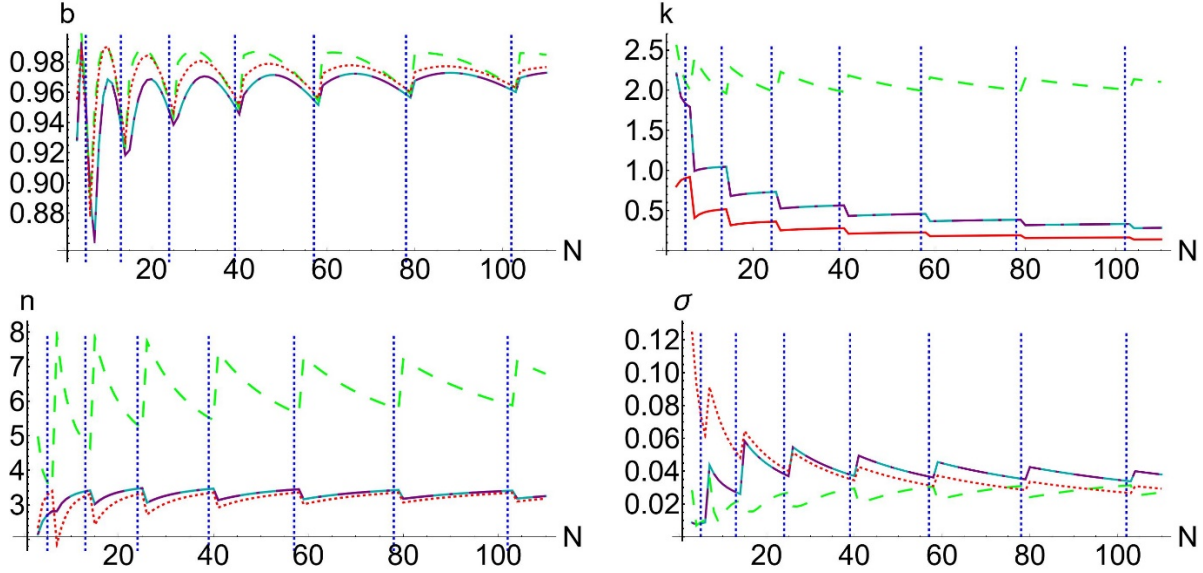

Fig. 10. *Values of the parameters of Hill fit for* the *probability to find solution for different functional dependences* between the *Householder angles and* the *register size N. The red dotted line corresponds to* $\omega = 2\pi - \varphi$, the *solid teal - to* $\omega = \pi$, the *dot-dashed purple - to* $\varphi = \pi$ *and* the *green dashed - to* $\omega = \varphi$. The top *left picture gives the parameters b and k correspondingly, and* the *bottom left and right pictures show the parameter n and the standard deviation of the fit* $\sigma$.

All parameters $b$, $k$, $n$ and the standard deviation have a periodic behavior depending on the number of iterations. We see that their maximum is close to the right side of the point where $k_{iter}$ change its value and slowly decrease until the end of the $k_{iter}$ value. Analytics shows that the maximal probability to find solution is achieved when $\omega = \phi$ as the plot of the parameter b shows that this functional dependence has maximal height. However, in the fit the local minima and maxima are smoothed, so b is not equal to the maximal probability to find solution. The parameters k and n are substantially larger for $\omega = \phi$ in comparison to when the other functional dependencies are used. This means that the robustness is highest for $\omega = \phi$ and the robustness in the other two cases is relatively close. The one when $\omega = \pi$ is slightly higher than the one when $\omega = 2\pi - \phi$. The error of the fit when $\omega = 2\pi - \phi$ is highest for small N, and decreases with increasing the register size.

Also, it is important to note two things: first, that $p(N, \varphi = \pi, \omega)$ and $p(N, \varphi, \omega = \pi)$, in both studied in previous sections (Sec 3.1. and Sec. 3.3.), phase matching modifications of one phase, have the same modified Hill fit. That is the reason why only $\omega = \pi$ is written in the *Table 1*. Second, when the latter phase matching modification is used, the parameters of the Hill fit are the same as in the Sec 3.1., however the values of the parameter omega are interchanged: $\omega = 2\pi - \varphi$ and $\omega = \varphi$.

Hill fit can be used to compare how the robustness changes with increasing the register size for different Grover's algorithm modifications. The easiest way to do that is to fit the numerical results for parameter $k$ for different register sizes (shown on *Fig. 10.*) in the best and the worst cases regarding the algorithm's stability when there are errors in the phase:

$$k_{OPH}^{BEST}(N) = 2.07982 - 0.950311 e^{-\varphi/3.11062} \tag{51}$$

$$k_{OPH}^{WORST}(N) = 0.180438 - 0.816018 e^{-\varphi/12.7592} \tag{52}$$

In the case of most robust and least robust cases we will give the fit of other two Hill function parameters $b$ an $n$ for different register size:

In the most stable and least stable cases, by using secondary fits, we obtain the corresponding expressions to the Hill function parameters b and n for different register size:

$$b_{OPH}^{BEST}(N) = \frac{e^{(N+57.1447)/14.5435}}{1 + e^{(N+57.1447)/14.5435}} - 0.0222395 \tag{53}$$

$$b_{OPH}^{WORST}(N) = \frac{e^{(N+25.7723)/7.8666}}{1 + e^{(N+25.7723)/7.8666}} - 0.0274633 \tag{54}$$

$$n_{OPH}^{BEST}(N) = 6.41228 + 3.49101 e^{-N/6.20068} \tag{55}$$

$$n_{OPH}^{WORST}(N) = 3.23431 + 0.615773 e^{-N/18.043} \tag{56}$$

As we will show, the results from the fits will allow for summarizing the large volume of simulation data and extracting the most important features. This will greatly simplify the analysis and predictions of the algorithm modifications' behavior for even larger N.

### 6.2. **Multiphase matching – only one gate in each iteration changes**

In this chapter, we will show some examples, where all phases $\omega_j$ are equal by modulo $\omega$, but their sign can differ. We will write the changing phase in the form $\omega_j = f(\omega, j)$ where j is the current iteration. Here, we also include remarks for the cases when φ changes instead of ω and $\varphi_j = f(\varphi, j)$. From now on, we will write phases in the form $\{\varphi_j, \omega_j\}$ for simplicity. In the next chapter, we will analyse the results.

#### 6.2.1. Alternate changing of the second phase ($ACSP$)

On *Fig. 11* are shown simulations of the probability to find solution when $\{\varphi_j, \omega_j\} = \{\varphi, (-1)^j \omega\}$. On the top left and right pictures are shown results of the simulation for register size 9 and 36, and on the bottom left and right - pictures for $N = 72$ and $N = 104$ respectively. The horizontal axis corresponds to φ, the vertical - to ω and the colour represents the probability to find solution.

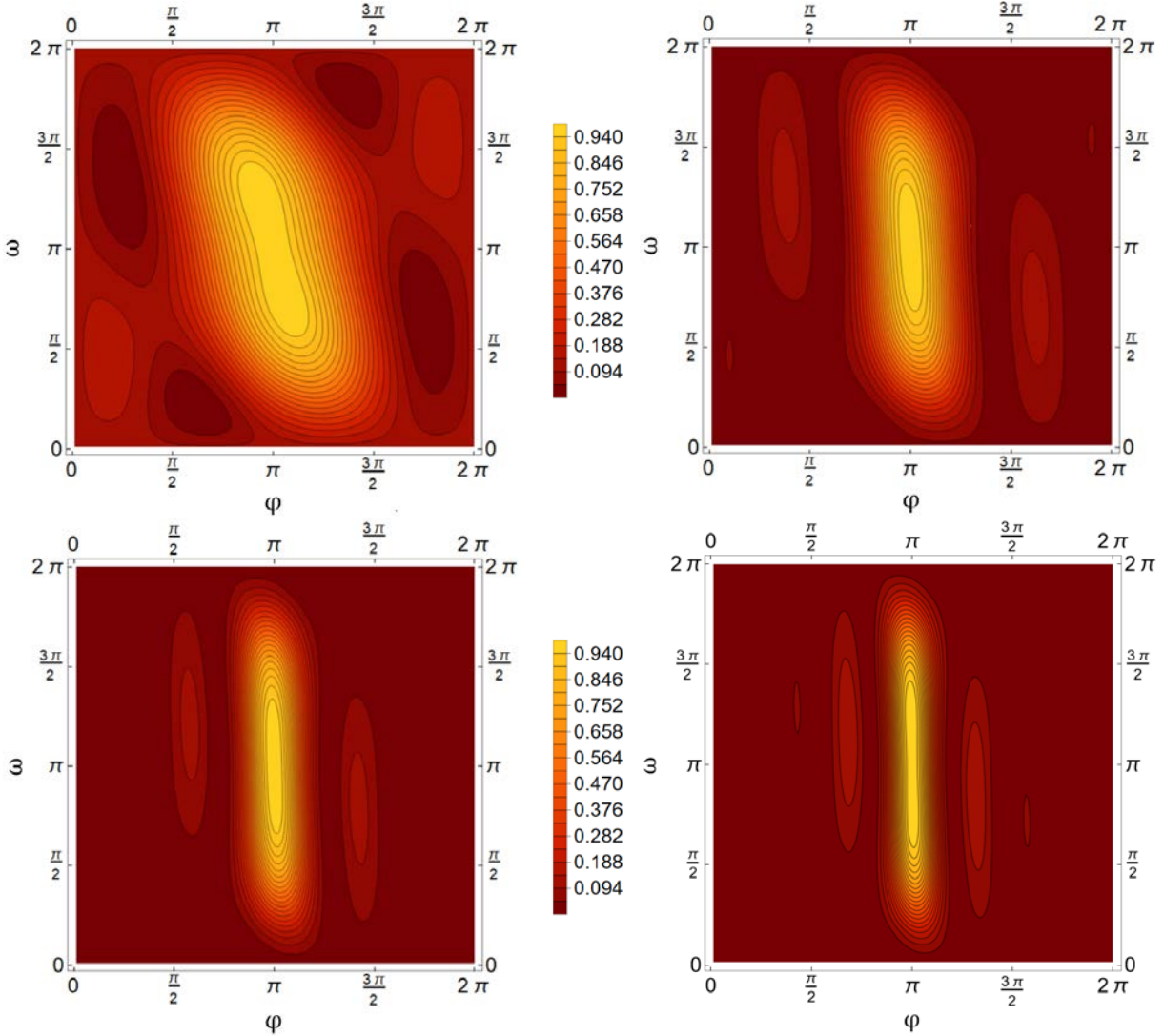

*Fig. 11. Probability to find solutions after $k_{iter}$ iterations of Grover's search algorithm with modification $\{\varphi_j, \omega_j\} = \{\varphi, (-1)^j \omega\}$ depending on the phases φ and ω. With yellow colour is depicted high and with red the low probability to find solution. The top left, top right, bottom left, bottom right pictures correspond to $N = 9$, $N = 36$, $N = 72$ and $N = 104$ respectively.*

Examples of few selected cross-sections of the figures above are shown in *Fig. 12*. On the horizontal axis is the angle φ and on the vertical - the probability to find solution. On each picture are shown simulations for different register size. The top left, top right, bottom left and bottom right figures depict register sizes 9, 36, 72 and 104 correspondingly. Each color represents different dependence between the phases $\varphi_j$ and $\omega_j$. More precisely: the green dashed line corresponds to $\omega = \varphi$, the red dotted line - to $\omega = 2\pi - \varphi$, the teal line - to $\omega = \pi$, the purple dot-dashed line - to $\varphi = \pi$.

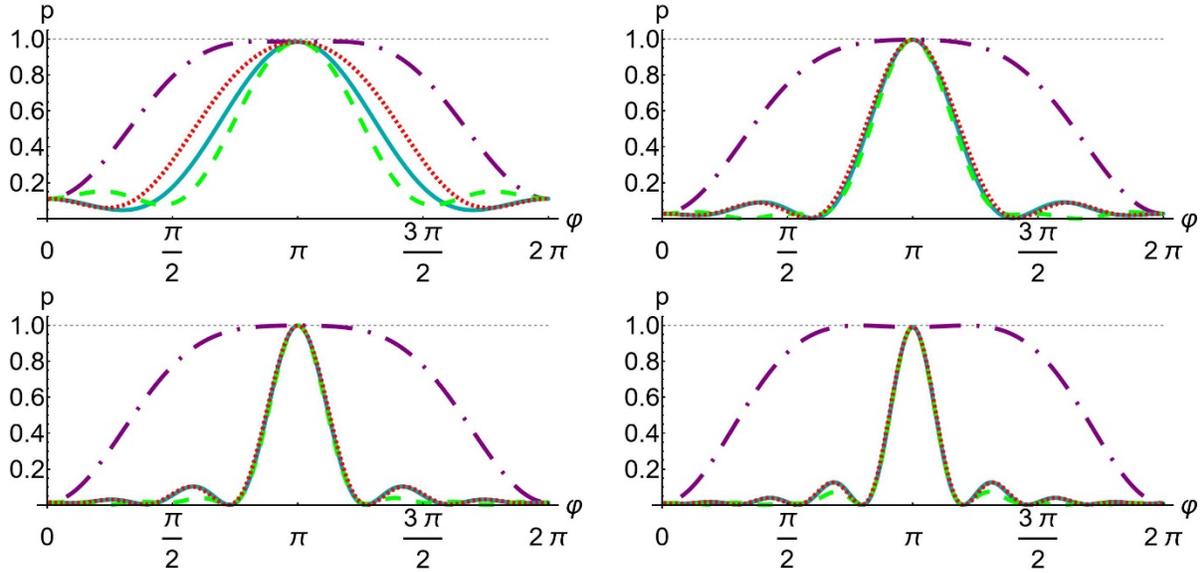

*Fig. 12. Probability to find solution p after $k_{iter}$ iterations of Grover's algorithm with angles used in gates construction given by $\{\varphi_j, \omega_j\} = \{\varphi, (-1)^j \omega\}$. From the shape of the different curves (each for a particular relation between the phases), the robustness of the corresponding algorithm implementation can be assessed. Different dashing and colour represent different functional dependence between the phases - the red dotted line corresponds to $\omega = 2\pi - \varphi$, the solid teal line to $\omega = \pi$, the dot-dashed purple line to $\varphi = \pi$ and the dashed green line to $\omega = \varphi$. The top left, top right, bottom left and bottom right pictures show the probability p for register sizes 9, 36, 72 and 104 respectively.*

For register sizes between 2 and 110, the cross-sections $\omega = \varphi$, $\varphi = \pi$, $\omega = 2\pi - \varphi$, $\omega = \pi$ were calculated. A modified Hill function was used to fit the respective bell like curves. On *Fig. 13* are shown the corresponding parameters of the fit. The top left, top right, bottom left and bottom right pictures show values of the parameters b, k, n and $\sigma$ respectively. The dashed green, dotted-red, solid teal and dot-dashed purple lines correspond to the following relations: $\omega = \varphi$, $\omega = 2\pi - \varphi$, $\omega = \pi$ and $\varphi = \pi$. The vertical blue dotted lines mark the register size when the required number of Grover's iterations increases by one.

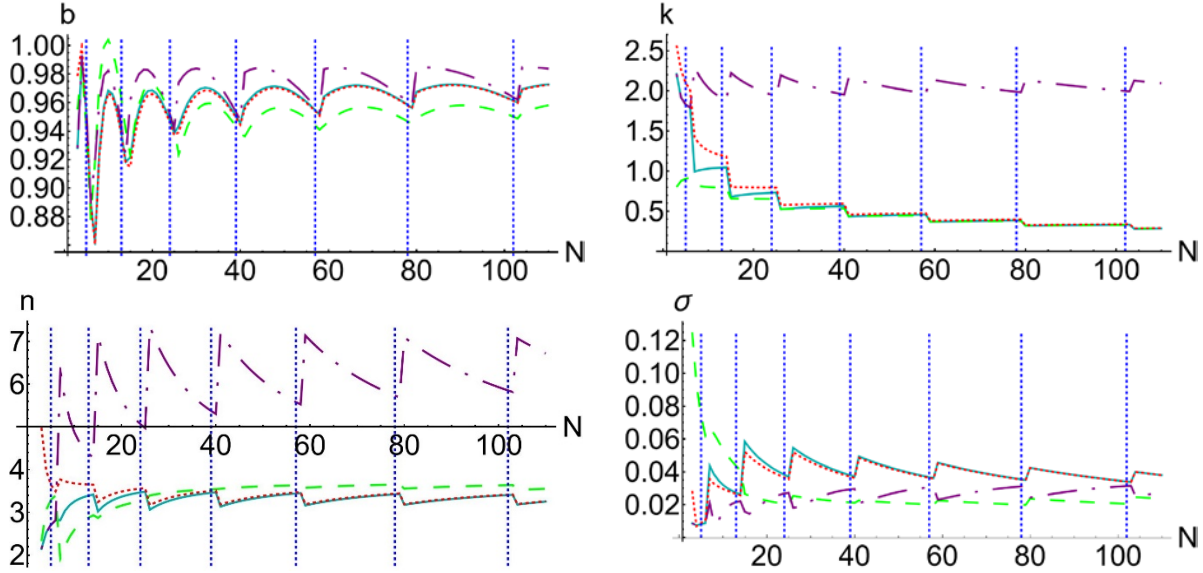

Fig. 13. Hill function parameter's values that describe the robustness of Grover's algorithm modification given by $\{\varphi_j, \omega_j\} = \{\varphi, (-1)^j \omega\}$, as a function of the register size N. Here, we give examples for four different functional dependences between the Householder reflection angles: $\omega = 2\pi - \varphi$ (depicted with red dotted line), $\omega = \pi$ (with solid teal line), $\varphi = \pi$ (with dot-dashed purple line) and green to $\omega = \varphi$ (with dashed green line). The top left, top right, bottom left and bottom right pictures give parameters b, k, n and the standard deviation respectively.

The modified Hill function fit can be used to compare characteristics of different cross-sections giving the probability $p$ in the $(\varphi_j, \omega_j)$ plane. In addition, by analysing how those parameters change with increasing the register size, we can make prognosis of their behaviour for large register size. As an example, from Fig. 13 we can predict that the maximal robustness (corresponding to the parameter k) is achieved when $\varphi = \pi$. On average, the robustness in this case remains constant with increasing the register size. The overall maximum of the probability to find solution (parameter b) is also maximal compared to the other functional dependences. The parameter n is larger in case of $\varphi = \pi$, compared to other cases. This means that in this case the probability decreases fastest at the end of the plateau. The robustness of the other three functional dependences decreases with increasing the register size in exponential way. After the interval with high probability to find solution, they decrease slower than the case of $\omega = \pi$, and their slope remains approximately constant with increasing the register size.

In order to approximately compare the robustness of this modification with the robustness of the other modifications, we can fit the obtained values of the parameters k in the best and worst cases and compare them with the corresponding quantities from the other cases studied in the previous sections of this work. Here, we compare the robustness for four different functional dependences between the angles ($\omega = \varphi$, $\omega = 2\pi - \varphi$, $\omega = \pi$ and $\varphi = \pi$). In this case the best and the worst functional dependences $\omega(\varphi)$ depend on the register size. As it increases, the best function dependence becomes closer to $\varphi = \pi$. This is the reason for us to use as best approximation (denoted by $k_{ACSP}^{BEST}(N)$) the relation $\varphi = \pi$ and as the worst approximation $\omega = \pi$ (denoted by $k_{ACSP}^{WORST}(N)$):

$$k_{ACSP}^{BEST}(N) = \frac{e^{(N+24.3169)/8.78275}}{1 + e^{(N+24.3169)/8.78275}} + 1.05907 \tag{57}$$

$$k_{ACSP}^{WORST}(N) = 0.280629 + 0.678234 e^{-N/34.1483} \tag{58}$$

The other two Hill functions' parameters $b$ and $n$ for the most and least robust cases are:

$$b_{ACSP}^{BEST}(N) = \frac{e^{(N+31.9503)/10.6365}}{1 + e^{(N+31.9503)/10.6365}} - 0.02367 \tag{59}$$

$$b_{ACSP}^{WORST}(N) = \frac{e^{(N-0.115542)/0.00716289}}{1 + e^{(N-0.115542)/0.00716289}} - 0.046407 \tag{60}$$

$$n_{ACSP}^{BEST}(N) = 6.28059 - 6.48707 e^{-N/6.37991} \tag{61}$$

$$n_{ACSP}^{WORST}(N) = \frac{e^{(N-19.7672)/5.2584}}{1 + e^{(N-19.7672)/5.2584}} + 2.59987 \tag{62}$$

For the same register size and the same number of solutions, there are other pairs of angles to be used in the j-th iteration $\{\varphi_j, \omega_j\}$ (with different dependences $\varphi_j(\varphi)$ and $\omega_j(\omega)$) that give the same, rotated (against the point $\{\varphi = \pi, \omega = \pi\}$) and/or mirrored (against $\phi = \pi$ axis) probability $p(\varphi, \omega)$. Some examples are:

1) $\{\varphi_j, \omega_j\} = \{\varphi, (-1)^{j+1}\omega\}$
   Coincides with $\{\varphi_j, \omega_j\} = \{\varphi, (-1)^j \omega\}$ when the axis $\omega$ is changed with $-\omega$.
   (Reflection against the line $\varphi = \pi$).
2) $\{\varphi_j, \omega_j\} = \{(-1)^j \varphi, \omega\}$
   Coincides with $\{\varphi_j, \omega_j\} = \{\varphi, (-1)^j \omega\}$ when the axes $\omega$ and $\varphi$ are exchanged.
   (Rotation against the point $\{\varphi = \pi, \omega = \pi\}$, on angle $\pi/2$).
3) $\{\varphi_j, \omega_j\} = \{(-1)^{j+1}\varphi, \omega\}$
   Coincides with $\{\varphi_j, \omega_j\} = \{\varphi, (-1)^j \omega\}$ after reflection against the line $\varphi = \pi$ and rotation on angle $\pi/2$ against the point $\{\varphi, \omega\}$.

### 6.2.2. Alternate changing of both first and second phase ($ACBP$)

On *Fig. 14* are shown simulations of the probability to find solution when $\{\varphi_j, \omega_j\} = \{(-1)^{j+1}\varphi, (-1)^j \omega\}$. The top left and right pictures show results for register sizes 9 and 36, and on the bottom left and right - the pictures for 72 and 104 respectively. Different colour represents different probability to find solution. The horizontal and vertical axes correspond to the angles $\phi$ and $\omega$.

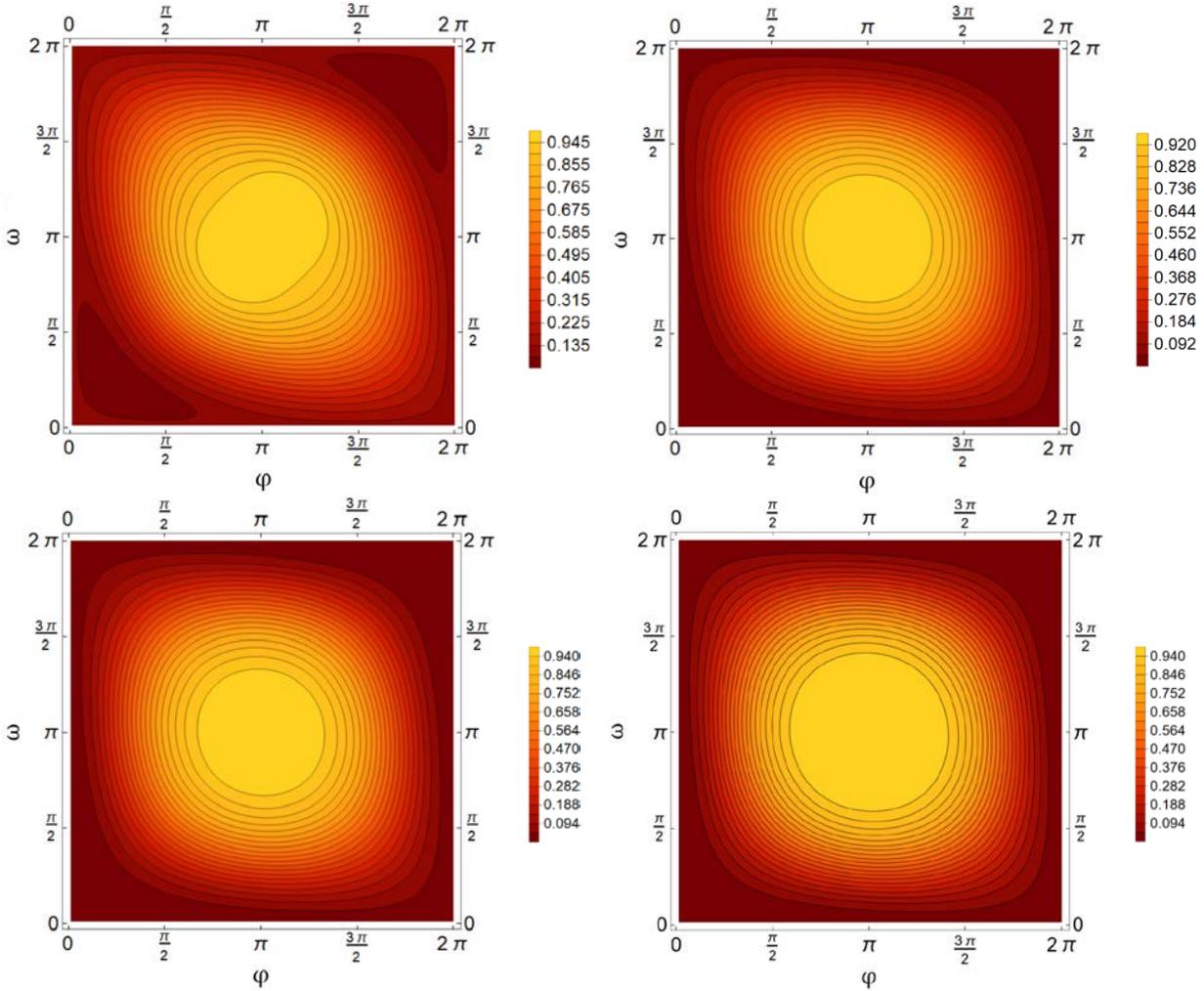

*Fig. 14. Probability to find solution of the latter modification of Grover's search algorithm depending on the phases φ and ω. Grover iteration phases are different each iteration and are calculated by $\{\varphi_j, \omega_j\} = \{(-1)^{j+1}\varphi, (-1)^j\omega\}$. The red colour shows low probability to find solution, and yellow - high. The top left, top right, bottom left, bottom right pictures give the results for register sizes $N = 9$, $N = 36$, $N = 72$ and $N = 104$ respectively.*

Here, on *Fig. 15*, we show the same cross sections as in the previous example. On the horizontal and vertical axis are given the angle $\phi$ and the probability to find solution. On each picture are shown the results for different register size – the top left, top right, bottom left and bottom right correspond to register sizes 9, 36, 72 and 104. Different functional dependences between the phases are represented by different color and dashing: the green dashed line corresponds to $\omega = \phi$, the red dotted line - to $\omega = 2\pi - \phi$, the teal line - to $\omega = \pi$, the purple dot-dashed line - to $\phi = \pi$.

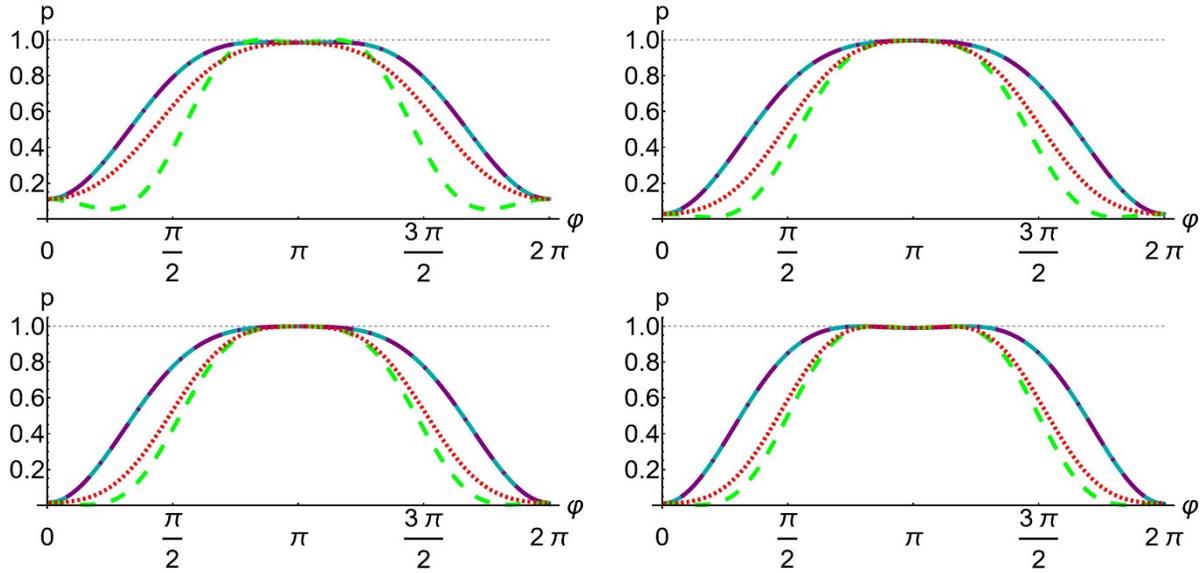

*Fig. 15. Probability to find solution of Grover's algorithm after required number iteration when each iteration angles are computed by $\{\varphi_j, \omega_j\} = \{(-1)^{j+1}\varphi, (-1)^j\omega\}$. Each coloured line corresponds different functional dependence between phases $\omega(\varphi)$. The dashed green, dot-dashed purple, solid teal, dotted red correspond to functional dependences $\omega = \varphi$, $\varphi = \pi$, $\omega = \pi$ and $\omega = 2\pi - \varphi$ respectively. On the horizontal axis is $\omega$ in case of $\varphi = \pi$ and $\varphi$ in all other cases, and p is on the vertical axis. On the top left, top right, bottom left and bottom right pictures are shown the probabilities to find solution for the following register sizes: 9, 36, 72 and 104 respectively.*

Such functions $p(\phi, \omega(\phi), N)$ with the studied relations $\omega(\phi)$ between its parameters for register sizes between 2 and 110 are fitted by using the modified Hill function. The fit parameters b, k, n and their standard deviation are shown on the top left, top right, bottom left and bottom right pictures of *Fig. 16*. Different colors indicate different functional dependences between phases The dashed green, dotted-red, solid teal and dot-dashed purple lines represent the relations $\omega = \phi$, $\omega = 2\pi - \phi$, $\omega = \pi$ and $\phi = \pi$.

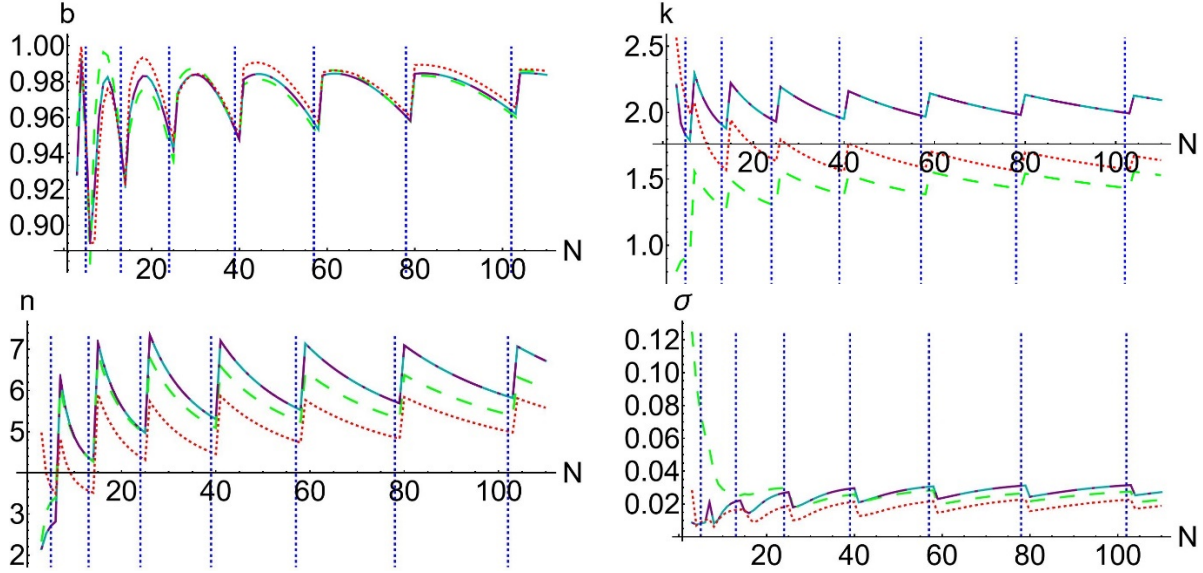

*Fig. 16. Hill fit parameter's values that describe the robustness of Grover's algorithm (modification $\{\varphi_j, \omega_j\} = \{(-1)^{j+1}\varphi, (-1)^j\omega\}$) as a function of the register size N. Top left, top right, bottom left and bottom right pictures give the parameters b, k, n and σ respectively. Different coloured and dashed lines correspond to differed functional dependences: $\omega = 2\pi - \varphi$ is depicted with red dotted line, $\omega = \pi$ - by solid teal line, $\varphi = \pi$ - by dot-dashed purple line and $\omega = \varphi$ - by dashed green line.*

From the results shown on *Fig. 16*, we can predict that the maximal robustness (parameter k) is achieved in two cases when $\omega = \pi$ and $\phi = \pi$. The robustness (on average) in those cases remains constant with increasing the register size. The overall maximum of the probability to find solution (parameter b) vary depending on the number of iterations needed. The overall larger value of the parameter n in the cases $\omega = \pi$ and $\phi = \pi$ shows that after the interval with high probability to find solution, the success probability $p(\phi, \omega(\phi), N)$ for those functions has a very steep slope. The robustness of the remaining two functional dependences overall also remains constant with increasing the register size. After the interval with high probability to find solution, the probability $p$ in those cases has less steep slope than the cases $\omega = \pi$ and $\phi = \pi$. On average their slope remains approximately constant with increasing the register size. This means that for large registers k(N) and n(N) are congruent and we can give an estimation of their value or large N.

Analogously to the previous chapter, here we also fit the obtained values of the parameter k in the best and worst cases and compare the stability with that of the other cases. Any other functional dependence between phases, will have parameter k (showing the robustness) close to the interval defined by $k_{ACBP}^{BEST}(N)$ and $k_{ACBP}^{WORST}(N)$:

$$k_{ACBP}^{BEST}(N) = \frac{e^{(N+24.3169)/8.78275}}{1 + e^{(N+24.3169)/8.78275}} + 1.05907 \qquad (63)$$

$$k_{ACBP}^{WORST}(N) = 1.46531 - 1.70115e^{-N/3.53768} \qquad (64)$$

In those cases, the other Hill function parameters have the following dependence of N:

$$b_{ACBP}^{BEST}(N) = \frac{e^{(N+31.9503)/10.6365}}{1 + e^{(N+31.9503)/10.6365}} - 0.02367 \qquad (65)$$

$$b_{ACBP}^{WORST}(N) = \frac{e^{(N+64.0767)/17.5393}}{1 + e^{(N+64.0767)/17.5393}} - 0.0238475 \qquad (66)$$

$$n_{ACBP}^{BEST}(N) = 6.28058 - 6.48707e^{-N/6.3799} \qquad (67)$$

$$n_{ACBP}^{WORST}(N) = 5.79722 - 6.73689e^{-N/4.40663} \qquad (68)$$

For the same register size and the same number of solutions, there are also functional dependences $\{\varphi_j, \omega_j\}$ that give the same results (or mirrored against the $\phi = \pi$ axis, or rotated against the point $\{\pi, \pi\}$). Three examples are:

1) $\{\varphi_j, \omega_j\} = \{(-1)^{j+1}\varphi, (-1)^{j+1}\omega\}$
   The same as $\{\varphi_j, \omega_j\} = \{(-1)^j\varphi, (-1)^j\omega\}$.
2) $\{\varphi_j, \omega_j\} = \{(-1)^{j+1}\varphi, (-1)^j\omega\}$
   Coincides with $\{\varphi_j, \omega_j\} = \{(-1)^j\varphi, (-1)^j\omega\}$ when mirrored against the line $\varphi = \pi$).
3) $\{\varphi_j, \omega_j\} = \{(-1)^j\varphi, (-1)^{j+1}\omega\}$
   Coincides with $\{\varphi_j, \omega_j\} = \{(-1)^j\varphi, (-1)^j\omega\}$ when mirrored against the line $\varphi = \pi$).

### 6.2.3. Each half of iterations with different first and second phases ($HIDP$)

Finally, we will study the robustness of a modification of Grover's algorithm in which the phases are $\{\varphi_j, \omega_j\} = \{(-1)^{[j/j_{max}]}\varphi, (-1)^{[j/j_{max}]}\omega\}$. On *Fig. 17* are shown examples for different register sizes that require even number of iterations: 9 (top left), 36 (top right), 72 (middle row left) and 104 (middle row right) respectively. Examples for register sizes that require odd number of iterations: 18 (bottom left), 51 (bottom right). Different colour represents different probability to find solution. The horizontal and vertical axes correspond to the angles $\phi$ and $\omega$.

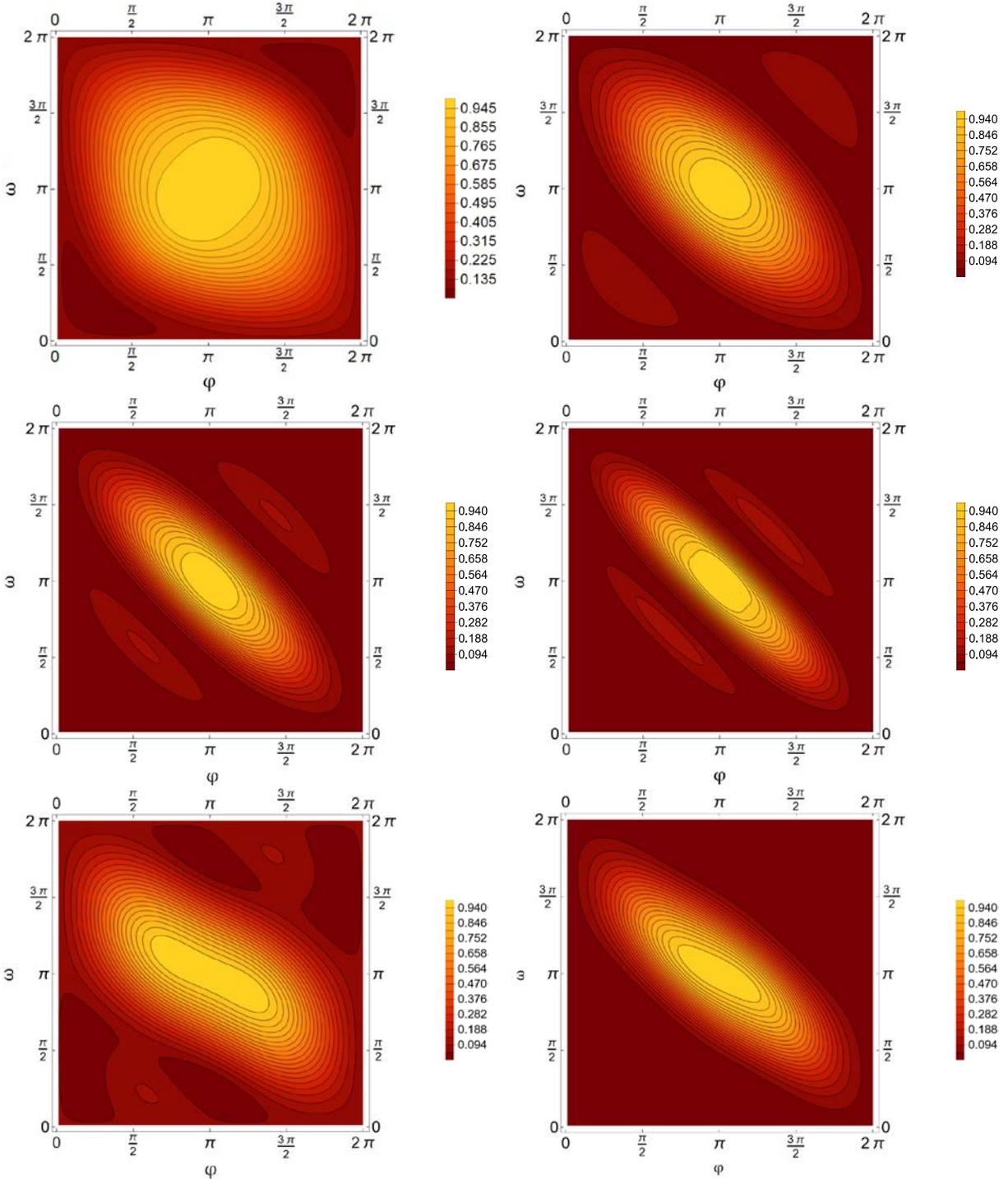

*Fig. 17. Probability to find solution of Grover's modification* $\{\varphi_j, \omega_j\} = \{(-1)^{\lfloor j/j_{max} \rfloor}\varphi, (-1)^{\lfloor j/j_{max} \rfloor}\omega\}$ *after the required number iterations as a function of the phases* φ *and* ω*. The yellow colour represents high probability to find solution and the red represents low probability. On the top left, top right, middle row left, middle row right, bottom left and bottom right pictures are shown simulations for register with sizes* $N = 9$, $N = 36$, $N = 72$, $N = 104$, $N = 18$ *and* $N = 51$ *respectively.*

Examples of some cross sections of the figure above are shown on *Fig. 18*. Each dashing and color represent a line with different dependence between the phases - the green dashed line corresponds to $\omega = \phi$, the red dotted line - to $\omega = 2\pi - \phi$, the teal line - to $\omega = \pi$, the purple dot-dashed line - to $\phi = \pi$. Each picture shows the results of simulations for different register sizes - the top left, top right, bottom left and bottom right correspond to register sizes 9, 36, 72 and 104 respectively.

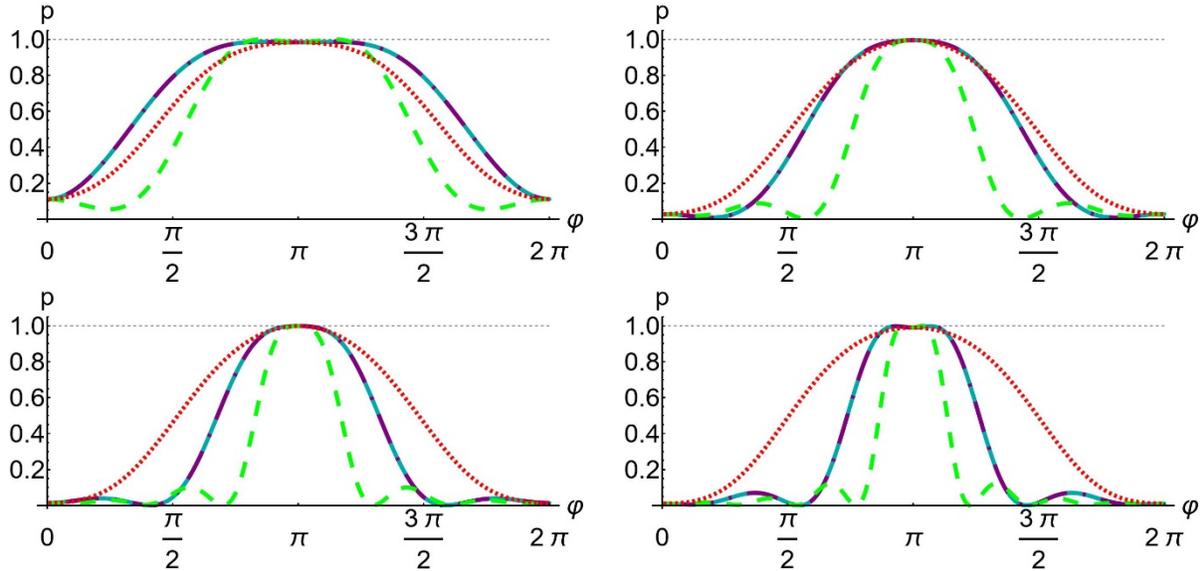

*Fig. 18. Probability to find solution of the Grover's algorithm in modification $\{\varphi_j, \omega_j\} = \{(-1)^{\lfloor j/j_{max}\rfloor}\varphi, (-1)^{\lfloor j/j_{max}\rfloor+1}\omega\}$ as a function of the dependence between the phases $\omega(\varphi)$. Each coloured line corresponds to different functional dependence of the errors in the phase. The dashed green, dot-dashed purple, solid teal, dotted red correspond to functional dependences $\omega = \varphi$, $\varphi = \pi$, $\omega = \pi$ and $\omega = 2\pi - \varphi$ respectively. On the horizontal axis is $\omega$ in case of $\varphi = \pi$ and $\varphi$ in all other cases, and the probability p is given on the vertical axis. Each picture represents the results for different register size: $N = 9$ on top left, $N = 36$ on top right, $N = 72$ on bottom left and $N = 104$ on bottom right.*

Similar to the investigated above $p(\varphi, \omega(\varphi), N)$ for all register sizes between 2 and 110 are fitted by using the modified Hill function on *Fig. 19*. The parameters of the fit and their standard deviation are shown on the top left, top right, bottom left and bottom right pictures. Each colour corresponds to different functional dependence between the phases. The dashed green, dotted-red, solid teal and dot-dashed purple lines are used for $\omega = \phi$, $\omega = 2\pi - \phi$, $\omega = \pi$ and $\phi = \pi$.

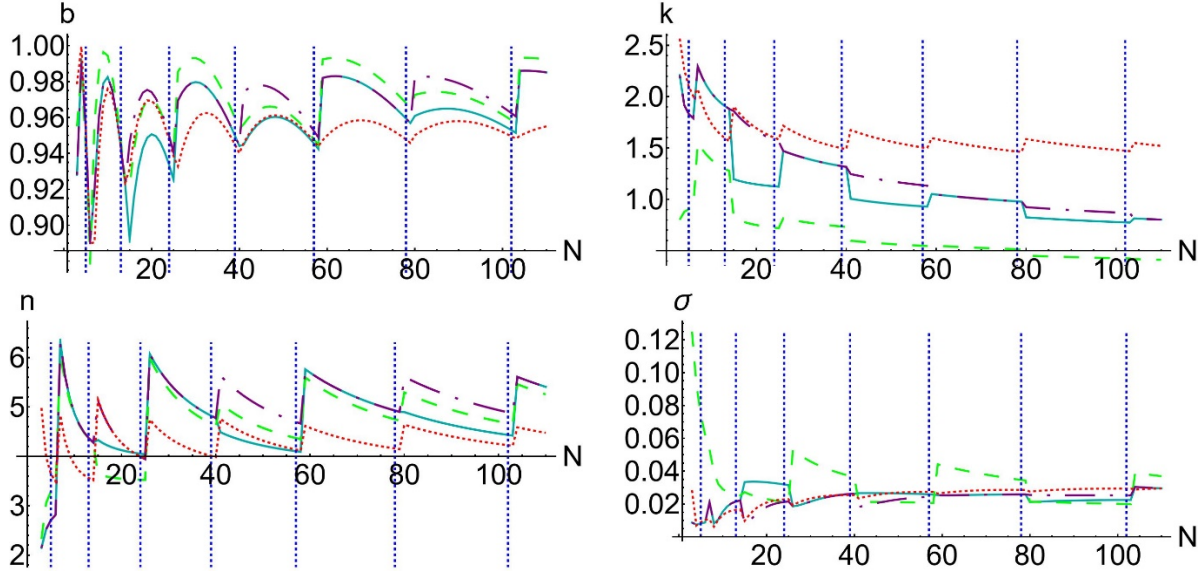

Fig. 19. Values of Hill fit's parameters of Grover's search algorithm modification $\{\varphi_j, \omega_j\} = \{(-1)^{\lfloor j/j_{max} \rfloor}\varphi, (-1)^{\lfloor j/j_{max} \rfloor}\omega\}$ as a function of the size of register N. The parameter b is shown on the top left, k – on top right, n – on the bottom left and σ - on the bottom right. Each coloured line corresponds to different functional dependence between the phases: $\omega = 2\pi - \varphi$ is depicted with red dotted line, $\omega = \pi$ - by with solid teal line, $\varphi = \pi$ - by dot-dashed purple line and $\omega = \varphi$ - by dashed green line.

From the results represented on *Fig. 19*, we can predict that the maximal robustness (corresponds to the parameter k) is achieved when $\omega = 2\pi - \phi$. The robustness in all cases (including in the best case) decreases exponentially with increasing the register size. In the most robust case, the overall value to the probability to find solution is the lowest (parameter b) compared with the other cases. In the most stable case the overall value of the parameter n is lowest – this means that outside the most robust interval the probability to find solution decreases relatively slowly. The slope of all functions remains approximately constant with increasing the register size.

In order to approximately compare the robustness (in the next chapter) of this modification with the robustness of the other modifications we can fit the obtained values of the parameter k in the best and worst cases and compare them with the robustness of the other cases. The fits of the functional dependences that give the best and the worst robustness for large registers are shown below:

$$k_{HIDP}^{BEST}(N) = 1.54396 - 1.26266 e^{-N/7.16425} \qquad (69)$$

$$k_{HIDP}^{WORST}(N) = 0.377259 - 0.928182 e^{-N/34.5457} \qquad (70)$$

The other Hill function parameters, that correspond to the fits with $k_{ACH}^{BEST}(N)$ and $k_{ACH}^{WORST}(N)$ are:

$$b_{HIDP}^{BEST}(N) = \frac{e^{(N+113.492)/20.3063}}{1 + e^{(N+113.492)/20.3063}} - 0.0456186 \qquad (71)$$

$$b_{HIDP}^{WORST}(N) = \frac{e^{(N+90.2055)/24.5754}}{1 + e^{(N+90.2055)/24.5754}} - 0.0237554 \qquad (72)$$

$$n_{HIDP}^{BEST}(N) = \frac{e^{(N+13.1927)/19.3789}}{1 + e^{(N+13.1927)/19.3789}} + 3.38676 \qquad (73)$$

$$n_{HIDP}^{WORST}(N) = 4.81735 - 13.3856 e^{-N/1.87989} \qquad (74)$$

Similarly, to the modifications shown in Sec. 6.2.1. and 6.2.2., here we will give some dependences $\varphi_j(\varphi, j)$ and $\omega_j(\omega, j)$ in order to obtain $\{\varphi_j, \omega_j\}$ that give the same results (up to reflection against line $\varphi = \pi$ and/or rotation against the point $\{\pi, \pi\}$). Three examples are:

1) $\{\varphi_j, \omega_j\} = \{(-1)^{[j/j_{max}]+1}\varphi, (-1)^{[j/j_{max}]+1}\omega\}$
   The same as $\{\varphi_j, \omega_j\} = \{(-1)^{[j/j_{max}]}\varphi, (-1)^{[j/j_{max}]}\omega\}$.
2) $\{\varphi_j, \omega_j\} = \{(-1)^{[j/j_{max}]+1}\varphi, (-1)^{[j/j_{max}]}\omega\}$
   Coincides with $\{\varphi_j, \omega_j\} = \{(-1)^{[j/j_{max}]}\varphi, (-1)^{[j/j_{max}]}\omega\}$ when rotated against the point $\{\varphi, \omega\}$, on angle $\pi/2$).
3) $\{\varphi_j, \omega_j\} = \{(-1)^{[j/j_{max}]}\varphi, (-1)^{[j/j_{max}]+1}\omega\}$
   The same as in the case of $\{\varphi_j, \omega_j\} = \{(-1)^{[j/j_{max}]+1}\varphi, (-1)^{[j/j_{max}]}\omega\}$.

## 7. Comparison of the robustness of the modifications

The obtained fits of the Hill function's parameters give the following advantages in studying the robustness of the reviewed modifications of the Grover's algorithm:

- Taking the average behavior of $b(N)$, $k(N)$ and $n(N)$ to remove the fluctuations of the Hills fit parameters for the same number of iterations but different register sizes. This gives us the overall behavior of $b$, $k$, $n$ and allows us to compare those parameters for the different modifications of the Grover's search algorithm.

- With increasing the register size, the "amplitude" of the fit's fluctuations for the same number of iterations of the fluctuation decreases. So, for sufficiently large registers those fluctuations can be neglected. Thus, using the smooth fits of the Hill's parameters will give us an accurate behavior and allows us to make prognosis of the values of the parameters of $b$, $k$ for $n$ even larger register sizes. We can calculate them for different functional dependences in the parameters and compare which of the modification is most robust overall.

- Allows us to approximately evaluate the robustness of the algorithm with different linear functional dependences between $\omega$ and $\phi$, that are not calculated in our examples.

On *Fig. 20* are shown the fits of Hill function's parameters for all studied modifications of Grover's search algorithm (with parameters $\{\varphi_j, \omega_j\}$). On the left are shown the cases where the modification is most robust against the error (with the appropriate functional dependences between the angles $\varphi$ and $\omega$) and on the right with functional dependences of the parameters where this modification is least robust. Each color represents the fit for different modification of the algorithm (the exact formulas of the fit can be seen written in the chapter of the corresponding modification). The phase matching algorithm's fit parameters are depicted by purple dot-dashed line – $k_{OPH}^{BEST}(N)$ (first row left), $k_{OPH}^{WORST}(N)$ (first row right), $b_{OPH}^{BEST}(N)$ (second row left), $b_{OPH}^{WORST}(N)$ (second row right), $n_{OPH}^{BEST}$ (third row left), $n_{OPH}^{WORST}$ (third row right). The parameters of Grover's algorithm modification with alternate changing of the second phase are depicted with solid teal line (equations for $k_{ACSP}^{BEST}(N)$, $k_{ACSP}^{WORST}(N)$, $b_{ACSP}^{BEST}(N)$, $b_{ACSP}^{WORST}(N)$, $n_{ACSP}^{BEST}(N)$ and $n_{ACSP}^{WORST}(N)$). The modification with both phases changing alternatively - with dashed green line (equations for $k_{ACBP}^{BEST}(N)$, $k_{ACBP}^{WORST}(N)$, $b_{ACBP}^{BEST}(N)$, $b_{ACBP}^{WORST}(N)$, $n_{ACBP}^{BEST}(N)$ and $n_{ACBP}^{WORST}(N)$). The last modification (where the first half of the iterations has the same operators and the second half has others) are shown in Sec 6.2.3. is depicted by dotted red line (equations for $k_{HIDP}^{BEST}(N)$, $k_{HIDP}^{WORST}(N)$, $b_{HIDP}^{BEST}(N)$, $b_{HIDP}^{WORST}(N)$, $n_{HIDP}^{BEST}(N)$ and $n_{HIDP}^{WORST}(N)$).

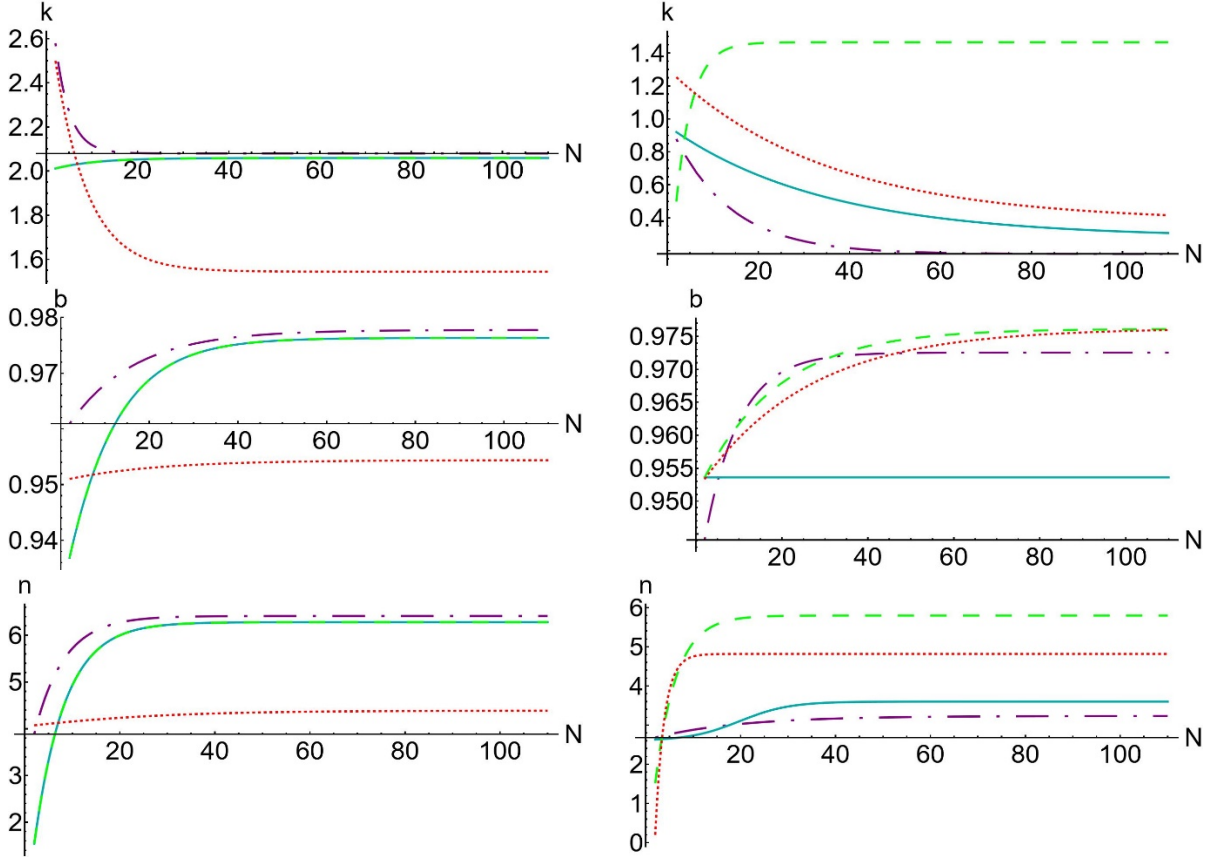

*Fig. 20. Fit of the Hill function approximation parameters for various modifications of Grover's algorithm as a function of the register size. The fits corresponding to the best functional dependence between the phases are shown on the left pictures and the worst - on the right ones. Different dashing and colouring represent different modifications. The purple dot-dashed line shows fits described by equations (51), (52), (53), (54), (55) and (56). The solid teal line shows fits described by equations (57), (58), (59), (60), (61) and (62). The green dashed line is used for the result of equations: (63), (64), (65), (66), (67) and (68). The dotted red line represents equations: (69), (70), (71), (72), (73) and (74).*

From the pictures in the first row of *Fig. 20*, it can be seen that the maximal robustness (with the optimal dependence between $\varphi$ and $\omega$) for large registers for $N = 100$ is approximately the same for almost all of the modifications. The fitting formulas can be used to obtain extrapolations for even larger register sizes (example for $N = 1000$) and the results confirm the above conclusion. The most robust modification (with its best functional dependence between the phases) is the one described in Sec. 6.1. with $k_{OPH}^{BEST}(1000) \cong 2.08$. The slightly less robust are the ones described in Sec. 6.2.1 and Sec. 6.2.2 $k_{ACSP}^{BEST}(1000) = k_{ACBP}^{BEST}(1000) \cong 2.05$. The least robust modification is the one shown in Sec. 6.2.3. In the latter the robustness is substantially worse $k_{HIDP}^{BEST}(1000) = 1.54$.

Of the modifications where the phase relations correspond to the worst robustness, the least stable is the one described in Sec. 6.1 with $k_{OPH}^{WORST}(1000) \cong 0.18$. The most stable modification is the one described in Sect. 6.2.2, with $k_{ACBP}^{WORST}(1000) \cong 1.46$. The stability of the two other modifications is $k_{ACSP}^{WORST}(1000) = 0.28$ and $k_{HIDP}^{WORST}(N) \cong 0.37$.

This shows that the most robust Grover's algorithm modification overall is the one described in Sec. 6.2.2 where the phases in each iteration are described by $\{\varphi_j, \omega_j\} = \{(-1)^{j+1}\varphi, (-1)^j\omega\}$.

The parameter b shows the overall height of the plateau of the Hill fit with width defined by the value of the parameter k. Its value depends on both- the modification and the functional dependence between the parameters. Here we will give values of the fit of $k(N)$ for all modifications studied in this work. This will allow us to obtain the evaluation of the probability to find solution for the corresponding modification when the functional dependence between the phases $\varphi$ and $\omega$. In the case of the overall most robust model shown in Sec. 6.2.2 the values of b in the best and worst functional dependences are $b_{ACBP}^{BEST}(1000) = 0.977$, $b_{ACBP}^{WORST}(1000) = 0.972$. And correspondingly for the least robust model shown in Sec. 6.1 the values of $b$ in best and worst cases are $b_{OPH}^{BEST}(1000) \cong b_{OPH}^{WORST}(1000) \cong 0.976$. The values of the parameter $b$ for the remaining cases studied are: the one in Sec. 6.2.1, namely $b_{ACSP}^{BEST}(1000) \cong 0.976$ and $b_{ACSP}^{WORST}(1000) \cong 0.953$, and the one in Sec.6.2.3, namely $b_{HIDP}^{BEST}(1000) \cong 0.954$ and $b_{HIDP}^{WORST}(1000) \cong 0.976$. This means that the intervals where $p(N, \varphi)$ has high probability have approximately the same high probability regardless of the relation $\omega(\varphi)$ and which of those modifications is chosen.

As expected, the value of the parameter n is higher in the cases when the value of k is higher. This gives the slope of the $p(N, \varphi)$ and fit of parameter n shows that the slope increases with increasing of N until it reaches its maximal value and then remain constant. This behavior is the same regardless of the modification or the dependence between the phases.

## 8. Conclusion

In this work we have studied the robustness of different modifications of Grover's search algorithm constructed by generalized Householder reflections to inaccuracies in their phases. We numerically calculate the various parameters of the one standard phase matching modification and three different multiphase matching modifications by using semi-empirical methods. Those parameters are robustness, average probability to find solution in the interval of high robustness and how fast the probability to find solution decreases outside this interval. In our calculations, we assume that there is a functional dependence between the phases in the generalized Householder reflections. Here, we study four different linear functional dependences between those phases. We compare these modifications to each other. By using extrapolations, we compare their characteristics for much larger registers. We show by numerical simulations that one of the multiphase matching modification (based on alternative flipping of the phase sign) achieve very high robustness against phase errors. This modification shows very high average probability to find solution (above 0.97) in this interval of high robustness.

## 9. Acknowledgments

This work was supported by the Bulgarian Science Fund under contract KP-06-M58/3 / 22.11.2021.